\documentclass[mathpazo]{cicp}

\usepackage[T1]{fontenc}
\usepackage[utf8]{inputenc}
\usepackage{ulem}
\usepackage[main=british, french, german, latin]{babel}

\usepackage{lineno,hyperref}
\usepackage{xcolor}

\modulolinenumbers[5]
\usepackage{amssymb}
\usepackage{amsmath}

\begin{document}
\title{Anomalous sorption kinetics of self-interacting particles by a spherical trap}

\author[C.~Astuto  et.~al]{Antonio Raudino \affil{1}, Antonio Grassi  \affil{1}, Giuseppe Lombardo \affil{1}, Giovanni Russo  \affil{2}, Clarissa Astuto \affil{3}\comma\corrauth, Mario Corti \affil{4}}
\address{\affilnum{1} Department of Chemical Sciences,  University of Catania, Viale A. Doria 6-95125, Catania, Italy\\ \affilnum{2}
Department of Mathematics, University of Catania, Viale A. Doria 6-95125 Catania, Italy \\
\affilnum{3} Applied Mathematics and Computational Science, King Abdullah University of Science and Technology (KAUST),  4700, Thuwal, Saudi Arabia \\
\affilnum{4} CNR-IPCF  Viale F. Stagno d'Alcontres, 37, 98158 Messina, Italy}
\email{{\tt clarissa.astuto@kaust.edu.sa} (C.~Astuto)}


\begin{abstract}
In this paper  {we propose a computational framework for the investigation of} the correlated motion between positive and negative ions exposed to the attraction of a bubble surface that mimics the (oscillating) cell
membrane. Specifically we aim to investigate the role of surface traps with substances freely diffusing around the cell. 
The physical system we want to model is an anchored gas drop submitted to a diffusive flow of charged surfactants (ions). When the diffusing
surfactants meet the surface of the bubble, they are reversibly
adsorbed and their local concentration is accurately measured.
The correlated diffusion of surfactants is described by a  {Poisson-Nernst-Planck (PNP) system, in which the drift term is given by the gradient of a potential which includes both the effect of the bubble and the Coulomb interaction between the carriers. The latter term is obtained from the solution of a self-consistent Poisson equation. 
For very short Debye lengths one can adopt the so called Quasi-Neutral limit which drastically simplifies the system, thus allowing for much faster numerical simulations. 
The paper has four main objectives.
The first one is to present a PNP model that describes ion charges in presence of a trap. The second one is to provide benchmark tests for the validation of simplified multiscale models under current development  {\cite{multiscale_mod}}. The third one is to explore the relevance of the term describing the interaction among  the apolar tails of the anions. The last one is to quantitatively explore the validity of the Quasi-Neutral limit by comparison with detailed numerical simulation for smaller and smaller Debye lengths.  
In order to reach these goals, we propose a simple and efficient Alternate Direction Implicit method  for the numerical solution of the non-linear PNP system, which guarantees second order accuracy both in space and time, without requiring solution of nonlinear equation at each time step. New semi-implicit scheme for a simplified PNP system near quasi neutrality is also proposed.} 
\end{abstract}


\maketitle

\section{Introduction}
The dynamical trapping of diffusing particles by either a single or a distribution of moving traps is an interesting topic that has been employed to model a variety of real problems in chemistry, physics and biology. Different ideal models have been proposed in the literature over the years. Most of the papers consider ideal traps where the diffusing particles impinging on its surface are irreversibly adsorbed (or chemically transformed). 
 
The broad field of biochemical reactions is grounded on the notion of stochastic encounters among diffusing particles. Although encounters do not guarantee chemical reactions among the colliding particles, they represent a key prerequisite for the reaction to occur. These concepts have been widely developed in the chemical-physics literature decades ago  \cite{Rice}. 
Since then, they have been extended to complex supra-molecular biochemical assemblies (like the protein searching for target sequences on DNA strands  \cite{RevModPhys8381}), to the prey-predator ecological models  \cite{variA,variB,variC,RevModPhys8381} or to the trapping phenomena in presence of static  \cite{RevModPhys85135} or oscillating fields \cite{REVELLI20041,PhysRevE.69.016105}. Trapping effects on the diffusive motion of particles are particularly relevant, introducing substantial deviations from the ideal behavior. Indeed, in normal diffusion the mean square displacement ${   \langle \textbf{r}^2 \rangle }$ of the diffusing particle is proportional to time, while
the trap modified behavior scales as: $\textbf{r}^2\approx Dt^\alpha$, where $D$ is the diffusion coefficient and $\alpha<1$ is the anomalous diffusion exponent (for a recent review see, e.g.,  {\cite{2012SMat89043S}}). 

In the previous examples, traps and preys have a comparable size, or, in other words, model focus on the very first event of catching a single pray (the so-called Mean First-Passage Time (MFPT) problems). There exists another broad class of traps (extended or multi-traps for short) and their dimension are much larger than that of a single prey. Large multi-traps act as scavengers for the impinging particles, adsorbing (reversibly or irreversibly) every particle reaching the interface. Typical examples are the growing crystals in a super-saturated solution, the chemical reactivity of a solid catalyst particle immersed in a sea of reactants, the nutrients diffusing toward the receptors-covered cell surface and so on. An important phenomenon occurring when considering large traps is that the catching history modifies the late catching efficiency through saturation of the available binding sites. Among the plethora of models describing trapping dynamics in presence of saturation effects, we would like to mention the classical Ward-Tordai model \cite{1946JChPh,borwankar,Liu2000}  explaining the diffusion-controlled coverage kinetics of a surface by a homogeneous distribution of ligands in solution. Another example is given by the chemoreception in a swimming cell, where molecules are adsorbed at the surface of a cell moving through a uniform distribution of ligands. This problem has been investigated several years ago by Wiegel \cite{WIEGEL1983283} and Berg \& Purcell \cite{BERG1977193} using different mathematical approaches.
 
As for the case of small traps, even large traps can be either immobile or diffusing by erratic motion or fluctuating around an equilibrium position. Particularly challenging is to predict the effect of motion on the capture efficiency of a trap, the surface of which oscillates by harmonic motion. The interest for these kinds of problems stems from the fact that all living cells experience active vibrations at their surface because of the large energy production associated to the intense cell internal metabolism \cite{Turlier2016}. 
Independently of the origin of the cell oscillations, they play a role in modulating the capture rate of ligands, messengers and nutrients from the outer space around the cell, because the advective contribution is much more effective than the diffusive one in transport phenomena. Such a claim is supported by the observation that surface motion may enhance heat \cite{vari4A,vari4B} or matter \cite{Xie2015} exchange at fluid interfaces. The increasing practical relevance of the wave-assisted transport devices (by ultrasound or microwaves) in many pharmaceutical and industrial fields share {s} common mechanisms based on the energy exchange between the adsorbates and the moving interfaces that act as scavengers of the adsorbed molecules. 

Motivated by these findings, we have recently undertaken a combined study by using theoretical, experimental and Molecular Dynamics approaches \cite{variRaudinoA,variRaudinoB,Raudino20168574,variRaudinoC,variRaudinoD,variRaudinoE,variRaudinoF,variRaudinoG}. Aim of this long-term project is to build-up a precise and tunable biomimetic system described by forced oscillating drops or bubbles, eventually dressed by a surfactant coat mimicking the cell membrane. Our goal is to investigate different aspects of the drop/bubble interfacial oscillations (frequencies, phases and amplitudes) and their modification upon the interactions with different surface-active agents. Although there exists a vast literature concerning the adsorption kinetics on the surface of oscillating drops and bubbles (a field often known as \textit{dilatation rheology} \cite{Miller2009}), our differential interferometric technique enables us to investigate oscillation amplitudes up to the sub-nanometric scale, highlighting new and still unexplored phenomena. 

In particular, the present work was motivated by one of our recent papers \cite{Raudino20168574} which reports the experimental trapping kinetics of a unidirectional diffusive flux of surfactants sticking at the surface of an oscillating gas bubble set in the middle of a diffusive flux. The unusual capture kinetics have been tentatively rationalized by a simple diffusion-based model coupled to oscillation-enhanced desorption. In the present paper, we extend the naive previous models by taking into account important improvements:
\begin{itemize}
	\item[A)]	A key prerequisite for trapping kinetics is the form of the potential energy. In addition to the particle-trap interaction (here modeled by a potential well of variable depth localized near the trap surface), in this paper we introduce interactions among the diffusing particles.  The inter-particle interactions are in general negligible in the bulk phase. However, they might become relevant when the particles are adsorbed on the trap where, depending on the strength of the potential, their concentration might reach high levels, even in the case of very dilute solutions. 
	\item[B)] We describe the diffusant as a fully dissociated species into univalent anions and cations and  {allow} for different sizes and chemical structure of anions and cations (that implies size-depending diffusion coefficients and specific ion-ion interactions) as well as for different strength of interactions between ions and the trap surface \cite{Poulichet5932,Breithbach2003}. 
    \item[C)] The governing equations are solved numerically with a controlled numerical error by a second order accurate method in space and time.
	This paper presents a novel version of the Alternate Direction Implicit (ADI) method, a second order accurate and stable scheme \cite{110002067615}. It is based on extrapolation technique and provides second order accuracy in space and time for the PNP system.
	\item[D)] The presence of a small parameter in the Poisson equation, the so called Debye length, poses a strong limitation on the time step. In order to overcome this difficulty a simplified single carrier PNP system is considered for which non conservative and conservative schemes are derived that are more stable than classical conservative ones.
\end{itemize} The plan of the paper is the following: in the next section we introduce the physical setup and the corresponding mathematical model. Section \ref{section_QNL} is devoted to the derivation of the quasi-neutral limit, which allows the treatment of the problem in the limit of vanishing Debye length.
Section \ref{section_simplified_PNP} is devoted to the development of new numerical methods, and to their test on a simplified PNP model. 
In Section \ref{section_space_discretization} we describe the discretization in space and in Section \ref{section_results} we apply the method developed in the previous section to the full PNP system, and analyze the numerical results.
Finally, in the last section we draw some conclusions.

\section{Physical setup and mathematical model} 
In this section we describe the experimental apparatus and the equations that govern the interaction of the carriers with the trap and between them.

\begin{figure}[tb]
	\begin{minipage}
		{.48\textwidth}
		\centering
		\includegraphics[width=\textwidth]{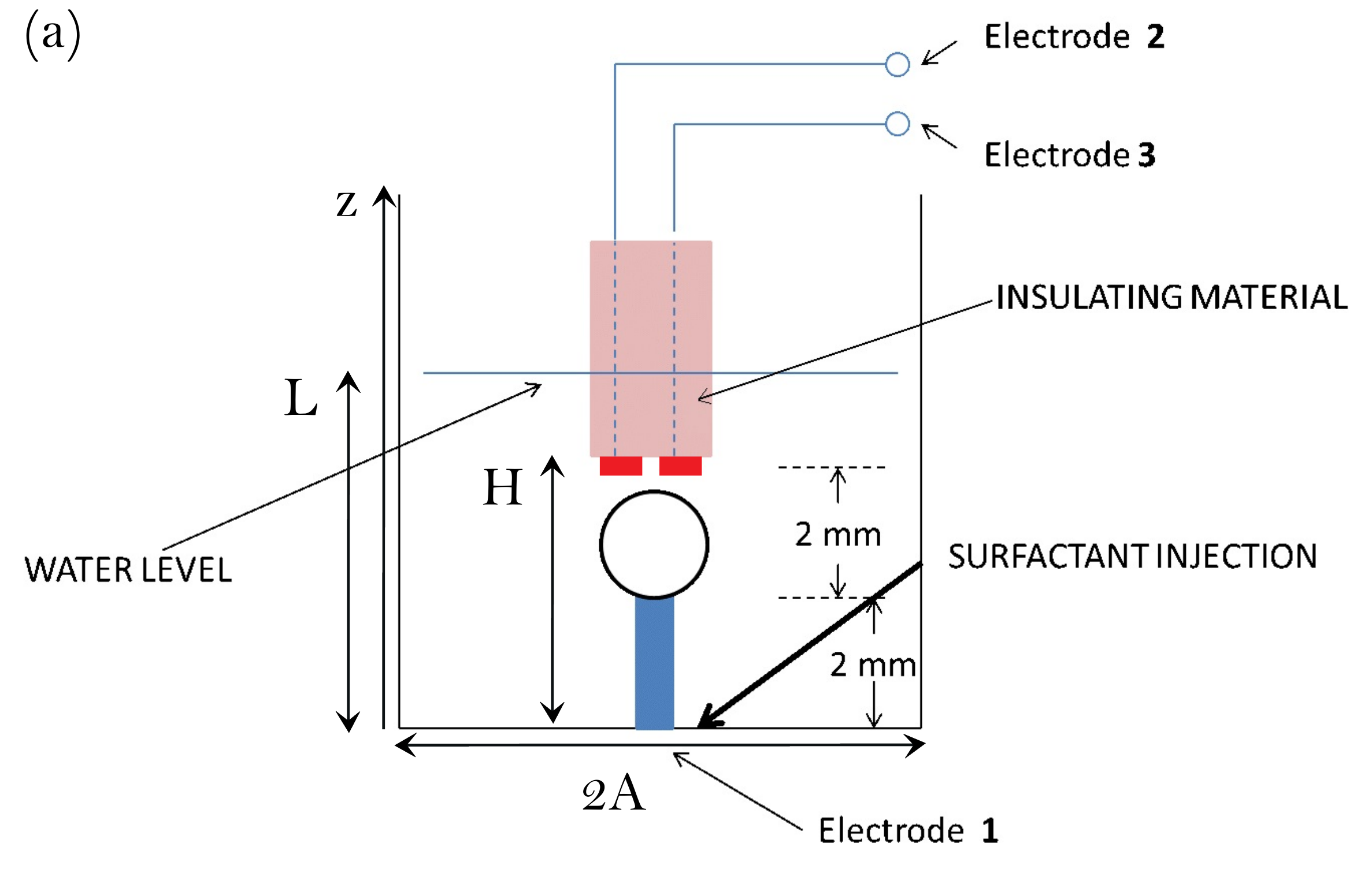}
	\end{minipage}
	\begin{minipage}
		{.48\textwidth}
		\centering
		\includegraphics[width=\textwidth]{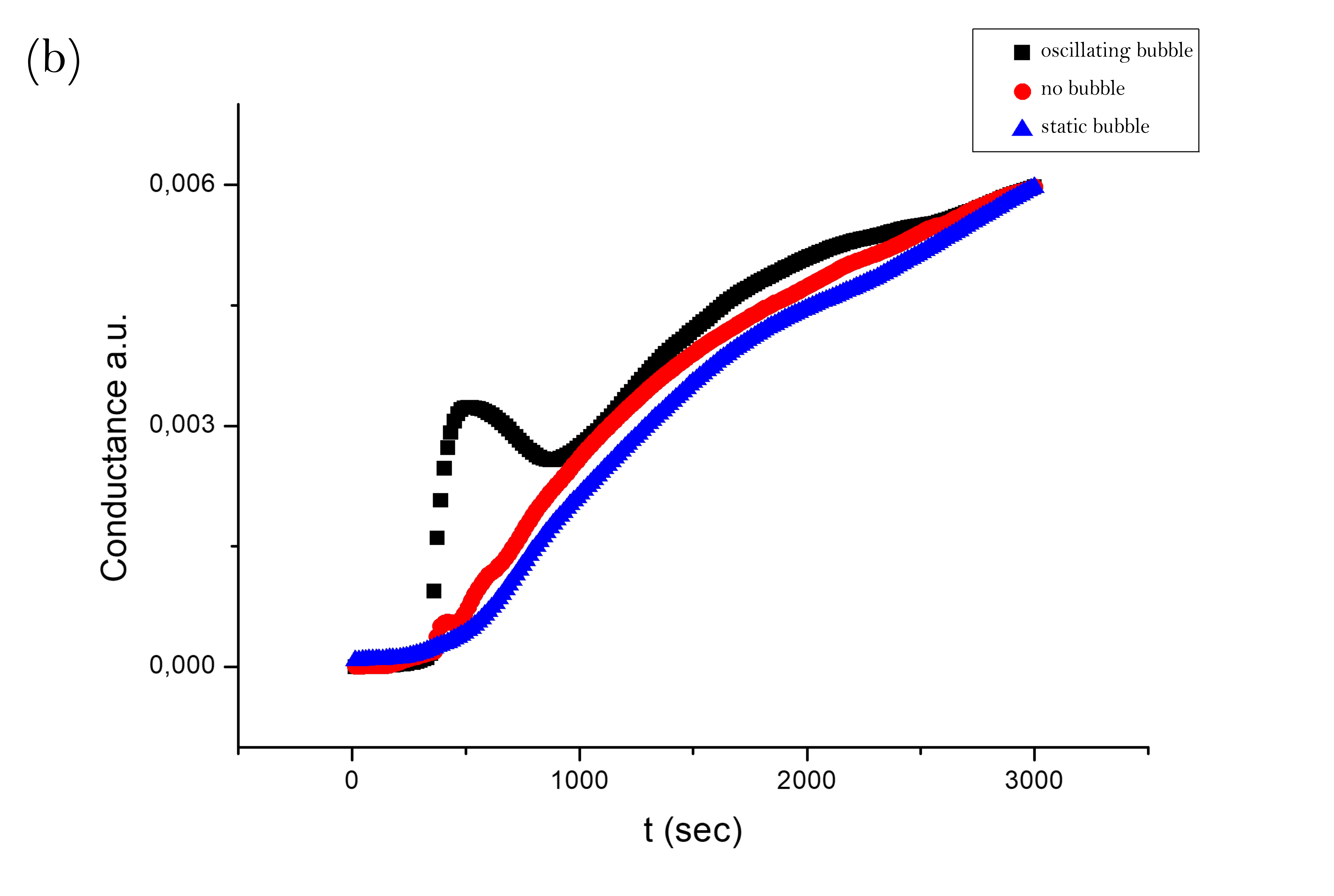}
	\end{minipage}
	\caption{\textit{Experimental domain and results. (a) Schematic setup of the real apparatus. The central sphere mimics the oscillating gas bubble. The detectors (in red) are located at distance $H$ from the bottom of the vessel. (b) Conductance of the aqueous solution measured over the bubble (electrodes 2 and 3 of left panel) versus time. Red line: no bubble, blue line: saturated bubble submitted to a flux of surfactants, black line: oscillating bubble submitted to a flux of surfactants (adapted from  \cite{Raudino20168574}). The lines thickness is an estimate of the experimental uncertainty of the conductivity measurements. }}
	\label{plot_setup}
\end{figure}

The experimental setup is shown in Fig.~\ref{plot_setup} (a)
and a typical plot describing the time evolution of the surviving diffusants in the neighborhood of the oscillating trap is shown in Fig.~\ref{plot_setup} (b).

The parameters employed in our numerical simulations closely match those used by us in our previous experiments \cite{Raudino20168574}. The bubble-containing vessel radius and height are both of order of a few millimeters, the bubble radius  is about $0.5\,mm$ and the  detector measuring the total ion flux past the oscillating bubble was set at a distance of $H$ from the center of the bubble (see Fig.~\ref{plot_setup} (a)).

The diffusion coefficients of  {the} ions were taken from the literature. In our experiments we mainly used Sodium Dodecyl Sulphate (\textit{SDS}) that in water solution fully dissociates  into a small sodium ion ($D^+ = {D}_{\rm{Na}^+}$, see  \cite{Lide}, and $D^- = D_{\rm SDS}$, see  \cite{ALSOUFI2012102,PMID20095711,PMID28623702}). An almost point-like distribution of the diffusants was set at $t=0$ near the bottom of the vessel ($z=0$). 


The correlated unsteady diffusion of self-interacting particles has been investigated over the years by a number of authors \cite{2015Entrp,Martzel_2001,MARIN20181903}. Because of its complexity from the theoretical and computational side, it is useful to introduce some approximation {s} that clarify the underlying physics of the coupled transport. 

\subsection{Diffusion in presence of a potential (trap). Correlation among the diffusants.}
Any statistical picture of transport phenomena in ionic solutions requires the calculation of the space and time evolution of the concentration of negative (anions) and positive (cations) species diffusing in a confined domain. Introducing the local concentrations of cations and anions: $c^+\equiv c^+(\textbf{r},t)$ and $c^-\equiv c^-(\textbf{r},t)$, continuity equation imposes that the time derivative of $c^{\pm\ }(\textbf{r},t)$ must equate the divergence of the ion flux: 
\begin{equation}
	\displaystyle \frac{\partial c^\pm\left(\textbf{r},t\right)}{\partial t}=-\nabla\cdot J^\pm(\textbf{r},t),
	\label{equation_flux}
\end{equation}
where $J^\pm(\textbf{r},t)$ denotes the particle current, under the action of a potential $V^\pm\left(\textbf{r},t\right)$.
              
Eq.~\eqref{equation_flux} can be separated into independent expressions for $J^+$ and $J^-$. Following the standard procedure developed in \ref{section_derivation} we obtain, in the dilute approximation:
\begin{eqnarray}
	\label{equation_definition_flux+}		
	J^\pm=\ -D^\pm\left(\nabla c^\pm+\ \frac{1}{k_BT}c^\pm\nabla V^\pm\right)
	\label{equation_flux_explicit}
\end{eqnarray}
where $D^\pm$ are the diffusion coefficients of positive and negative ions (assumed to be constant throughout the whole system),  $k_B$ is the Boltzmann constant, $T$ is the absolute environment temperature (assumed to be constant) and $\nabla$ stands for the gradient. The interaction potentials $V^\pm$ experienced by the positive and negative diffusing particles may depend also on $c^{\pm}$. Without loss of generality, they can be partitioned as: 
\begin{eqnarray}
	V^\pm=\ V_{\rm{ion-bubble}}^\pm+\ V_{\rm{ion-ion}}^\pm                 	       
	\label{equation_definition_ext}
\end{eqnarray}
where the term $V_{\rm{ion-bubble}}^\pm$ describes the interaction between a specific ion, located at a generic position $\textbf{r}$, and the (possibly oscillating) interface, while the term $V_{\rm{ion-ion}}^\pm$  accounts for the interaction among the diffusing ions. The main contribution of the ion-ion interactions is the electrostatic term, written as: $V_{\rm{ion-ion}}^\pm=\ \pm Z^\pm q\varphi(\textbf{r})$, where  $Z^\pm$ is the number of unit charges of the ions (in the present study $Z^+=\ Z^-=1$), $q$ is the (absolute) electron charge and $\varphi\left(\textbf{r}\right)$ is the still unknown electrostatic potential among the ions. In the present study, an additional term has been introduced in order to consider the hydrophobic pairing among the apolar tails of the anionic molecules. In the simplest mean-field picture, the attraction among the hydrocarbon tails of the anions can be described by a term proportional to the concentration $c^-$ of the anions, thus we write:
\begin{eqnarray}
	\label{equation_beta_p}
	V_{\rm{ion-ion}}^-= {-q\varphi-\beta c^-},\quad
	V_{\rm{ion-ion}}^+= +q\varphi
\end{eqnarray}
where the parameter $\beta$ measures the strength of the (attractive) Van der Waals interactions among the apolar tails of the bulky anions ($\beta$ is of order of some $k_BT$ units), also known as steric effect, and it is strictly related to the tail length or bulk. Of course, no hydrophobic interaction occurs among the small hydrophilic cations, for this reason this term does not appear in the expression for $V^+_{\rm{ion-ion}}$ describing the cation-cation interactions. From a physical standpoint, the bulky anions experience two opposite forces: a repulsive one due to the electrostatic repulsion among them, and an attractive one due to the Van der Waals attraction among the bulky tails. In other words, the physical picture of a solution of surfactants is different from the one of an ideal plasma becasue here particles with the same charge (anions) feel also an attractive potential. This peculiar term introduces considerable deviations between the system investigated by us and the one of an ideal plasma like an electrolyte solution. Combining the above results, we get from Eqs.~(\ref{equation_flux}-\ref{equation_beta_p}) a set of two coupled Nernst-Planck (NP) equations \cite{LU20106979,FPEHannes1996,2005nfpebook} valid for dilute solutions:
\begin{eqnarray}
	\label{equation_cphi+}\displaystyle \frac{\partial c^{+}}{\partial t} &=& D^+\left(\Delta c^+ + \frac{1}{k_BT}\nabla\cdot\left(c^+\nabla\left(V^+_{\rm{ion-bubble}}+q\varphi\right)\right)\right)\\ \label{equation_cphi-}
	\displaystyle \frac{\partial c^{-}}{\partial t} &=& D^-\left(\Delta c^- + \frac{1}{k_BT}\nabla\cdot\left(c^-\nabla\left(V^-_{\rm{ion-bubble}}-q\varphi - \beta c^-\right)\right)\right)	
\end{eqnarray}
The system reflects conservation of mass and describes the influence of concentration gradients and electric field on the flux of diffusing chemical species \cite{JUNGEL200083,doi:10.1080/03605309908821456}, specifically ions. 

To close  {Eqs}.~(\ref{equation_cphi+}-\ref{equation_cphi-}) we need an equation for the electrostatic potential $\varphi$, which is given by the Poisson equation relating the potential to the ion charge as follows
\begin{equation}
	\displaystyle -\epsilon_0\epsilon_r{\Delta\varphi}=q\left(n^+-n^-\right)
	\label{equation_poisson_phi}
\end{equation}
where $\epsilon_0$ is the vacuum permittivity, $\epsilon_r$ is the relative permittivity ($\epsilon_r = 78$ in water) and $n^\pm$ are the ion charge density which are proportional to the ion concentrations $c^\pm$ by the relation:
\begin{eqnarray}
	n^\pm=c^\pm\frac{N_A\rho^\pm}{\tilde{m}^\pm} 
	\label{equation_n_pm}
\end{eqnarray}
where $N_A$ is the Avogadro's number, $\tilde{m}^\pm$ the molecular mass of ions (expressed in Kg/mol) and $\rho^\pm$ their mass densities (in $\rm{Kg}/m^3$). After simple rearrangement of Eq.~\eqref{equation_poisson_phi} (see \ref{section_dimensionless}) we rewrite the system as 
\begin{eqnarray}
	\label{equation_c_U_eps+}\displaystyle \frac{\partial c^{\varepsilon,+}}{\partial t} &=& D^+\left(\Delta c^{\varepsilon,+} + \nabla\cdot\left(c^{\varepsilon,+}\nabla\left(U^+_{\rm{ion-bubble}}+U^\varepsilon\right)\right)\right)\\	 \label{equation_c_U_eps-}\displaystyle \frac{\partial c^{\varepsilon,-}}{\partial t} &=& D^-\left(\Delta c^{\varepsilon,-} + \nabla\cdot\left(c^{\varepsilon,-}\nabla\left(U^-_{\rm{ion-bubble}}-U^\varepsilon-\frac{\beta}{k_BT} c^{\varepsilon,-}\right)\right)\right) \qquad\\ 
	\label{equation_lap_Ueps}-\varepsilon^2\Delta U^\varepsilon &=& \frac{c^{\varepsilon,+}}{m^+}-\frac{c^{\varepsilon,-}}{m^-}
\end{eqnarray}
where $\displaystyle \varepsilon=K^{-1/2}$ and, setting $\rho:=\rho^+ = \rho^-\approx\rho(H_2O)$, the constant $K$ turns out to be $ \displaystyle K=\frac{q^2N_A\rho}{\epsilon_0\epsilon_rk_BT m_0}$ (see Table~\ref{table_parameters} for the values of the parameters and the related discussion in Section \ref{section_dimensionless}). The constant $\varepsilon$ is related to the Debye length that is of the order of nanometers for concentrations $c^\pm \approx 10^{-6}$, while the device and the bubble lengths are of the order of millimeters. This may justify the quasi-neutrality approximation we propose in this paper. 

From the Poisson equation, concentrations are functions of $ \varepsilon$ hence we define $\displaystyle c^{\varepsilon,\pm} := c^\pm(\varepsilon)$ in Eq.~(\ref{equation_c_U_eps+}-\ref{equation_lap_Ueps}) to underline the presence of stiffness also in the PNP equations and the direct consequences in the definition of the time step.

The concentrations of particles $c^{\varepsilon,\pm}$ satisfy the Nernst-Planck Eqs.~(\ref{equation_c_U_eps+}-\ref{equation_c_U_eps-}) coupled self-consistently to the Poisson Eq.~\eqref{equation_lap_Ueps} for the electrostatic potential $U^\varepsilon$ (with $\displaystyle U^\varepsilon := \frac{q\varphi}{k_BT}$, see \ref{section_dimensionless}).

 {System}~(\ref{equation_c_U_eps+}-\ref{equation_lap_Ueps}) is defined in a bounded domain $\Omega\subset \mathbb{R}^3$ and it is completed with an initial condition for the concentrations and zero boundary conditions for the flux because particles cannot enter or leave the domain. The boundary condition for the potential $U^\varepsilon$ derives by the zero normal electric field along the wall because it is  {made of} an insulating material.
\begin{equation}
		\label{equation_BC_IC}	              	
		c^{\varepsilon,\pm}(t=0)=c_{0}^{\varepsilon,\pm}(\textbf{r})   \quad  {\text{ in } \Omega}
\end{equation}
\begin{equation}
\label{equation_BC_IC2}
\displaystyle J^\pm\cdot\hat{n} = 0 \quad  {\text{ on } \partial\Omega}, \quad \frac{\partial U^\varepsilon}{\partial \hat{n}} = 0  \quad  {\text{ on } \partial \Omega}	
\end{equation}
where $\hat{n}$ is the outgoing normal unit vector of $\partial\Omega$. 

\begin{figure}[tb]
	\begin{minipage}
		{.48\textwidth}
		\centering
		\includegraphics[width=\textwidth]{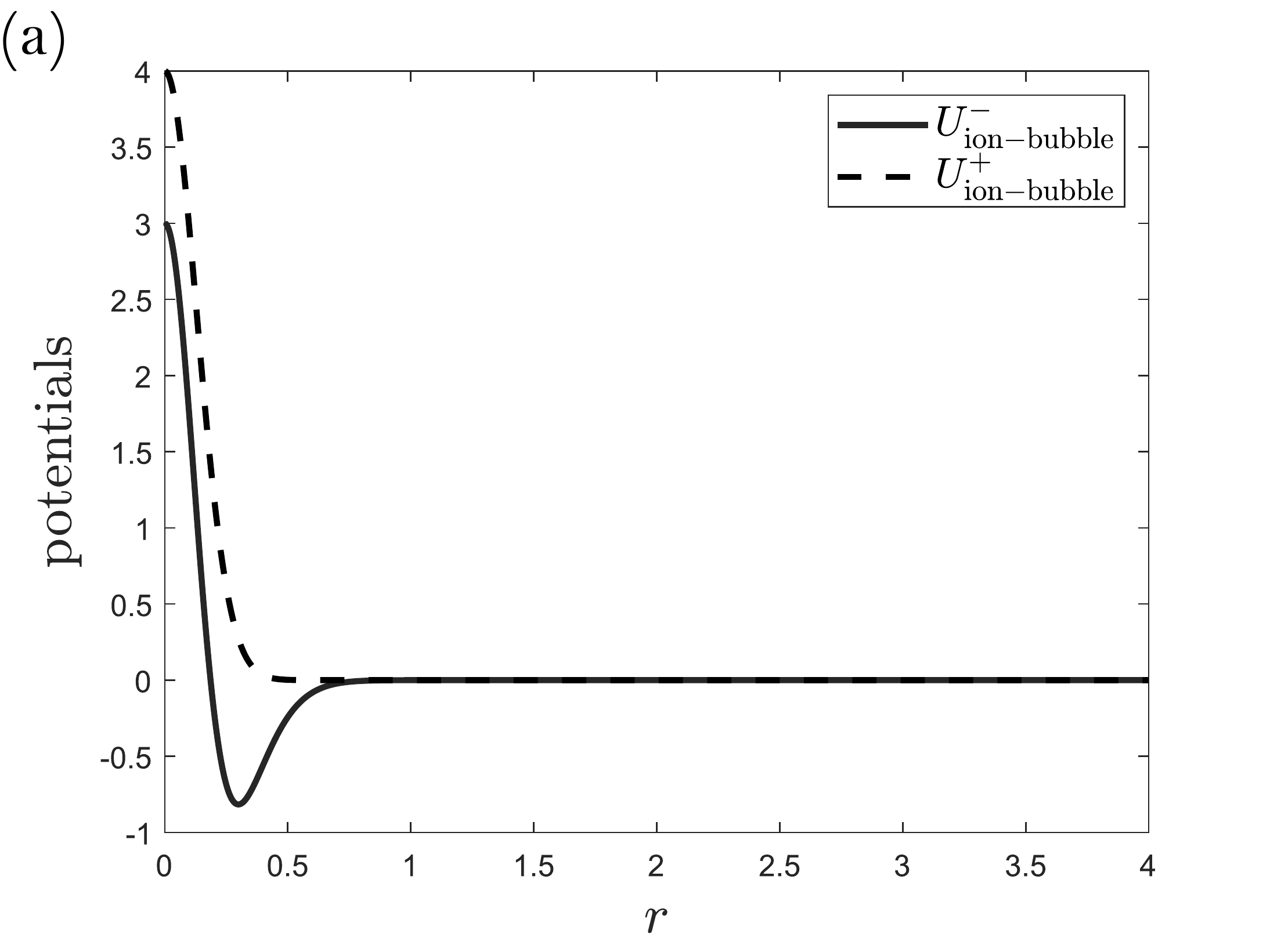}
	\end{minipage}
	\begin{minipage}
		{.48\textwidth}
		\includegraphics[width=\textwidth]{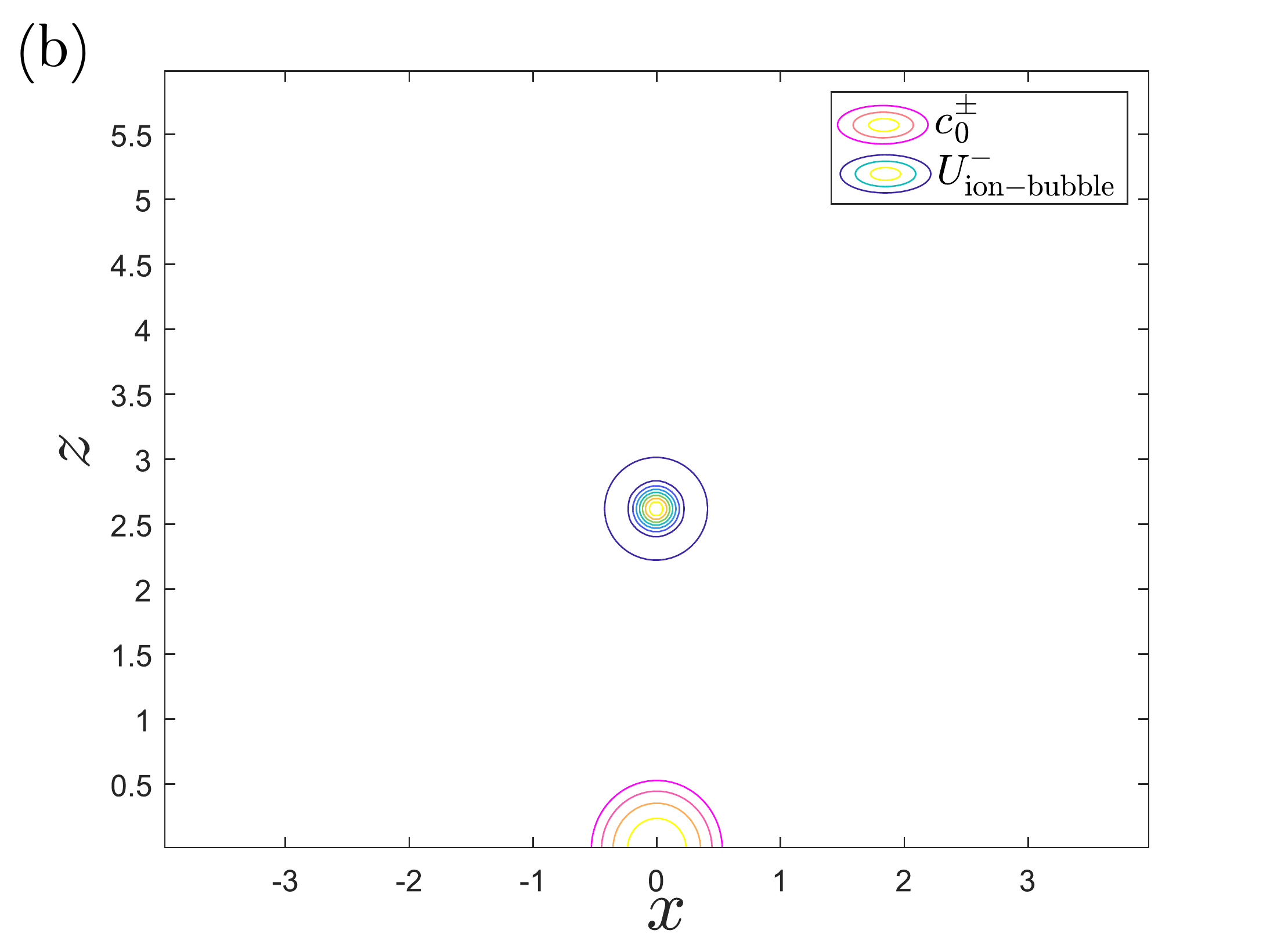}
	\end{minipage}
	\caption{\textit{Ingredients of the model: external potentials and initial condition. (a) Representation of the potentials $U_{\rm{ion-bubble}}^\pm$ in 1D as a function of $r$ at $z = z_c$, center of the bubble and reported in Table~\ref{table_parameters}. The solid line is the potential for the anions and the dashed line is the potential for the cations. (b) Contour plots in $(x,z)$ plane, $y=0$, of the anions potential $U^-_{\rm{ion-bubble}}$ representing the bubble (center of the panel) and of the initial condition at the bottom of the vessel. The potential fixes the diffusants concentration in the aqueous phase at $t = 0$.}}
	\label{QNL_ext_plot}
\end{figure}

Adopting axisymmetric cylindrical coordinates $r$ and $z$,
the expression for the initial condition we use in our computations is the following
\begin{eqnarray}
	c^{\varepsilon,\pm}(t=0,r,z)&=c_{0}^{\varepsilon,\pm}(r,z)=\frac{2v_0^\pm}{\left(2\pi\sigma^2\right)^{3/2}}\exp\left(-\left(r^2+z^2\right)/2\sigma^2\right)\quad 
	\sigma \in \mathbb{R}
	\label{equation_IC}
\end{eqnarray}
where $v_0^{\pm}$ denote the total volume of positive and negative ions. Total charge neutrality and the assumption that
positive and negative ions have the same mass density, impose that $v_0^+/m^+ = v_0^-/m^-$ (see Fig.~\ref{QNL_ext_plot} (b)). The analytical form of the initial condition \eqref{equation_IC} has been selected to mimic the experimental conditions where a small amount of surfactant was injected at $t=0$ by a micro syringe near the bottom of the vessel ($z=0$). Numerical solutions of the set of Eqs.~(\ref{equation_c_U_eps+}-\ref{equation_c_U_eps-}) submitted to the boundary and initial conditions in Eqs.~(\ref{equation_BC_IC}-\ref{equation_IC}) yield the expected concentrations $c^{\varepsilon,\pm}$ as a function of space and time.

In the specific experiments we are going to discuss, we measure negative and positive ions reaching a detector placed at a distance $H$ apart from the bottom (see Fig.~\ref{plot_setup} (a)). The key ingredient to describe trapping phenomena is $U^\pm:=U^\pm\left(\textbf{r},t\right)$, a still undefined potential (the so-called chemical potential) that acts on the diffusing species. The structure of $U^\pm_{\rm{ion-bubble}}$ is, in general, rather intricate, depending upon the particle-bubble interaction near its surface (that for an oscillating bubble is time-dependent) and upon the interaction among the charged particles.

The interfacial forces are very sensitive to the chemical nature of the ions and of the interface (see e.g.,  \cite{CHANG1995}). In this paper we investigate the important case were anions and cations have a specific chemical structure and cannot be approximated as point-like charges. In particular, the investigated anions bear a long apolar tail (i.e., they are surfactants). This peculiar mixed structure gives rise to strong interfacial hydrophobic interactions. As a result, surfactants accumulate at the air-water interface with the hydrophobic tails protruding toward the gas phase. Conversely, small cations (that balance the negative charge of anions) are hydrophilic and tend to escape from the air-water interface in order to be surrounded by water molecules. 

Here we focus on hydrophobic anions (surfactants) electrically balanced by small hydrophilic cations. These asymmetric systems are representative of wide classes of chemical/biochemical compounds (detergents, lipids, emulsifiers). We use the  approximation that the concentrations $c^{\varepsilon,\pm}$ of the diffusing particles are low (of order of $10^{-6}$ moles/liter in the experiments  {in \cite{Raudino20168574}}), so their mutual interaction is negligible (beside the electrostatic one which is taken into account). Near the interface, however, their concentration is higher and their mutual interactions could play a significant role. This effect is subject of current investigation. 

A convenient phenomenological potential for the anion-bubble interaction $U_{\rm{ion-bubble}}^-$ as a function of $R=\sqrt{r^2 + (z-z_c)^2}$, where $(0,z_c)$ denotes the center of the bubble, takes the form: 
\begin{eqnarray}
	\label{equation_Ue+} \displaystyle 	U_{\rm{ion-bubble}}^+\left(\textbf{r}\right)&=&a_1\exp(-b_1R^2)\\ 
	\label{equation_Ue-} \displaystyle 	U_{\rm{ion-bubble}}^-(\textbf{r})&=&a_2\exp(-b_2R^2)-a_3\exp(-b_3R^2)\\
	\nonumber z_c \in \Omega,&&  a_1,a_2,a_3,b_1,b_2,b_3\in\ \mathbb{R}
\end{eqnarray}
A typical shape of the anion/cation potential is reported in Fig.~\ref{QNL_ext_plot} (a). The impenetrability of the bubble with positive and negative ions is modeled by a repulsive potential (described by the first terms in Eqs.~(\ref{equation_Ue+}-\ref{equation_Ue-})). The second and negative term in Eq.~\eqref{equation_Ue-} describes the favorable hydrophobic interactions between the long tail of the anionic surfactant and the air/water interface set at the bubble boundary. As previously said, the attractive energy contribution contained in Eqs.~(\ref{equation_Ue+}-\ref{equation_Ue-}) is applied to the hydrophobic anions while cations-surface attraction is automatically modeled by the electrostatic potential. 
\section{Quasi-Neutral Limit (QNL)}
\label{section_QNL}
Considering Eq.~\eqref{equation_lap_Ueps} we notice that the term $\varepsilon^2$ is very small, thus making the problem stiff. If we try to solve the coupled Poisson-Nernst-Planck (PNP) system \cite{EISENBERG2007,LU20112475} with a fractional step scheme then the computational cost would be prohibitively high, for two reasons. First stability requirements impose a strong restriction for the time step size reducing the efficiency of the solution procedure. Secondly the Poisson equation requires a computation of a linear system of order $N^2\times N^2$ (if we pose the number of points of the discretization $N :=  N_r=N_z$) that we cannot avoid. For these reasons we consider the approach based on the so called \textit{Quasi-Neutral Limit} (QNL), as seen, for instance, in \cite{jungel}. The model consists of continuity equations for ions and a Poisson equation for the electrostatic potential in a bounded domain. 
In \cite{jungel} it has been shown that both species diffuse at the same rate with a common
diffusivity that is intermediate between the ones of the two species. 

To apply the quasi-neutral limit we need to perform the limit $\varepsilon\rightarrow 0$ in the Eqs.~(\ref{equation_c_U_eps+}-\ref{equation_lap_Ueps}). For simplicity of notation we include the term $\beta c^{\varepsilon,-}/(k_BT) $ in  $U_{\rm{ion-bubble}}^-$.
Dividing Eqs.~(\ref{equation_c_U_eps+}-\ref{equation_c_U_eps-}) by $m^+$ and $m^-$ respectively and defining $\displaystyle C^{\varepsilon,\pm} := \frac{c^{\varepsilon,\pm}}{m^\pm}$ we have
\begin{eqnarray}
	\label{QNL-system+}
	\displaystyle \frac{\partial C^{\varepsilon,+}}{\partial t}&=&D^+( {\Delta}C^{\varepsilon,+}+ {\nabla}\cdot\left(C^{\varepsilon,+} {\nabla}U_{\rm{ion-bubble}}^+\right)+ {\nabla}\cdot\left(C^{\varepsilon,+} {\nabla}U^\varepsilon\right)) \\ 
	\label{QNL-system-}
	\displaystyle
	\frac{\partial C^{\varepsilon,-}}{\partial t}&=&D^-\left( {\Delta}C^{\varepsilon,-}+ {\nabla}\cdot\left(C^{\varepsilon,-} {\nabla}U_{\rm{ion-bubble}}^-\right)- {\nabla}\cdot\left(C^{\varepsilon,-} {\nabla}U^\varepsilon\right)\right) \quad\\ 
	\label{QNL-system}
	\displaystyle
	-\varepsilon^2 {\Delta}U^\varepsilon&=&\left(C^{\varepsilon,+}-C^{\varepsilon,-}\right)
\end{eqnarray}
In the limit $\varepsilon \rightarrow 0 $ we obtain $C^{\varepsilon,+} = C^{\varepsilon,-}$, so we pose 
\begin{eqnarray}
		C :=   C^{\varepsilon,+} = C^{\varepsilon,-}.	
\end{eqnarray}
Adding and subtracting Eqs.~(\ref{QNL-system+}-\ref{QNL-system-}) leads to
\begin{eqnarray}
	\label{eq_sum}		\displaystyle 2\frac{\partial C}{\partial t}&=&\left(D^++D^-\right) {\Delta C}+D^+ {\nabla}\cdot\left(C {\nabla}U_{\rm{ion-bubble}}^+\right) \\ \nonumber
	&&+D^- {\nabla}\cdot\left(C {\nabla}U_{\rm{ion-bubble}}^-\right) +\left(D^+-D^-\right) {\nabla}\cdot\left(C {\nabla}U^\varepsilon\right) \\
	\label{eq_diff}
	\displaystyle 0&=&\left(D^+-D^-\right) {\Delta C}+D^+ {\nabla}\cdot\left(C {\nabla}U_{\rm{ion-bubble}}^+\right)\\  \nonumber
	&& -D^- {\nabla}\cdot\left(C {\nabla}U_{\rm{ion-bubble}}^-\right)+\left(D^++D^-\right) {\nabla}\cdot\left(C {\nabla}U^\varepsilon\right)
\end{eqnarray}
Solving Eq.~\eqref{eq_diff} for $\nabla\cdot\left(C {\nabla}U^\varepsilon\right) $ and replacing its expression in Eq.~\eqref{eq_sum} we can eliminate the drift term coming from the electrostatic interaction and the final expression is
\begin{eqnarray}	
		\displaystyle \frac{\partial C}{\partial t}=D_{\rm{eff}}{\Delta C}+D_{\rm{eff}}{\nabla}\cdot\left(C {\nabla}U_{\rm{ion-bubble}}\right)
	\label{limit}
\end{eqnarray}
where the '{effective}' diffusion coefficient is the harmonic mean of the diffusion coefficients of the two ion species,  $D_{\rm{eff}} ={2D^+D^-}/{(D^+ + D^-)}$, 
and the effective potential is the sum of the two ion-bubble potentials:
$U_{\rm{ion-bubble}} = (U_{\rm{ion-bubble}}^+ + U_{\rm{ion-bubble}}^-)$.

We assume that the initial condition is well prepared, i.e. it is compatible with \textit{local} charge neutrality
\begin{equation}
	C^+(t=0) = C^-(t=0) =: C_0 
	\label{QNL-IC}
\end{equation}

\section{Efficient methods for a simplified PNP model}
\label{section_simplified_PNP}
In the first part of this section we present a simplified PNP model, which allows us to test two new numerical schemes. 
The first one is a variant of the classical \textit{Alternating Direction Implicit} (ADI) method which allows second order accuracy for the non-linear PNP system, and which is only linearly implicit, therefore it does not require solution of nonlinear equations. 
The second scheme is aimed at solving the stiffness problem arising for small (but non negligible) Debye lentghts. 

\subsection{A simplified PNP model}
A simplified model is obtained by assuming that the concentration of the carrier density of the anions is a known function of space and time. We shall denote by $c_{\rm A}(x,y,t)$ the known background concentration, and assume for simplicity that the domain is a 2D square $\Omega$.
We denote simply by $c(x,y,t)$ the unknown concentration, and assume that the only potential is the electrostatic potential due to the interaction with the background. 
The resulting equations for the simplified model are therefore: 

\begin{eqnarray} \label{eq_ADI}
	c_t   &=& \nabla\cdot (c\nabla\phi ) + \mu \Delta c \\
	-\varepsilon \Delta \phi &=& c - c_{\rm A}  \\ 
	\label{eq_ADI_in}
	c(x,y,t=0) &= & c_{\rm in }(x,y)
\end{eqnarray}  
where $c_{\rm A} \equiv c_{\rm A}(x,y,t)$ and $\mu$ is the diffusion coefficient.
For simplicity we consider $\Omega = [-\pi,\pi]^2$, with periodic boundary conditions both on
$c$ and $\phi$.
Eq.~\eqref{eq_ADI} may be rewritten in a separate form as 
\begin{equation}
	c_t = L_x(\phi_x)c + L_y(\phi_y)c
	\label{eq:c_t2}
\end{equation}
where  {$L_x(\phi_x)c \equiv (\phi_x c)_x + \mu c_{xx}$, $L_y(\phi_y)c \equiv (\phi_y c)_y + \mu c_{yy}$}, and using standard notation, the subscript denotes partial derivative on $\phi$ and $c$.
\subsection{Second order ADI discretization.}
A standard ADI discretization 
for the model PNP system can be described as follows.

Given $c^n(x,y)\approx c(x,y,t^n)$ we solve  {system}~(\ref{eq_ADI}-\ref{eq_ADI_in}) as
\begin{eqnarray}
	\label{eq_standard_ADI}
	-\Delta \phi^{n} &=& \frac{c^{n} - c_{\rm A}}{\varepsilon }\\ \label{eq_standard_ADI_2}
	\tilde{c} &=& c^n + \frac{\Delta t}{2}L_y\left(\phi_y^{n}\right)\tilde{c} + \frac{\Delta t}{2}L_x\left(\phi_x^{n}\right)c^n \\
	\label{eq_standard_ADI_3}
	{c}^{n+1} &=& \tilde{c} + \frac{\Delta t}{2}L_y\left(\phi_y^{n}\right)\tilde{c} + \frac{\Delta t}{2}L_x\left(\phi_x^{n}\right)c^{n+1} 
\end{eqnarray}
We assume the equation is discretized on a regular square Cartesian grid. The Poisson equation is solved  {by a} Fourier spectral method. A conservative finite difference space and time discretization, which ensures exact mass conservation for both ions (within round-off errors) and second order accuracy in space,  {has been used}. Both drift and diffusion terms in Eqs.~(\ref{equation_c_U_eps+}-\ref{equation_lap_Ueps}),(\ref{QNL-system+}-\ref{QNL-system}),\eqref{limit} are discretized by central difference, making sure that the mesh P\'{e}clet number \cite{Wesseling2023600} is always within the stability threshold.

The presence of the term $\phi^n$ in Eqs.(\ref{eq_standard_ADI_2}-\ref{eq_standard_ADI_3}) prevents one from obtaining second order accuracy in time.

Several techniques can be adopted to obtain a second order method in space and time, still avoiding fully implicit solvers. A general technique to construct linearly implicit second and high order methods for a wide class of evolutionary partial differential equations is based on a suitable use of IMEX (IMplicit-EXplicit) schemes \cite{IMEX}. These methods are based on identifying the terms of the system which are responsible for the stiffness and treating them implicitly with the IMEX machinery. 

In  {this} specific context, however, it is possible to use simpler alternatives.
In this section we propose a simple technique that provides second order accuracy in time, still avoiding the implicit computation of the nonlinear term. 

An alternative to the general IMEX approach consists in computing a predicted value solving Eqs.~\eqref{eq_standard_ADI} and \eqref{eq_standard_ADI_2}, updating the potential to $\tilde{\phi}$ by solving the Poisson equation with $\tilde{c}$ in place of $c^n$, and then adopting again Eqs.~\eqref{eq_standard_ADI} and \eqref{eq_standard_ADI_2} with $\phi^n$ replaced by $\tilde{\phi}$.
Let us denote this method as the \textit{standard second order scheme}.
The cost of a full time step with this method is almost double than the standard ADI step.
An even simpler alternative is obtained by extrapolating the concentration, and here we describe this strategy. 

Given the concentration at time $t^n$ and $t^{n-1}$, we extrapolate the concentration at time $t^{n+1/2}$
\begin{eqnarray}
\label{eq_extrapol}	\displaystyle c^{n+1/2} = \frac{3}{2} c^n - \frac{1}{2} c^{n-1}  
\end{eqnarray}
then compute $\phi^{n+1/2}$ by 
solving the Poisson equation 
\[
    -\Delta \phi^{n+1/2} = \frac{c^{n+1/2} - c_{\rm A}(x,y,t^{n+1/2})}{ \varepsilon},
\]
and apply the ADI method to solve the Eqs.~(\ref{eq_ADI}-\ref{eq_ADI_in}) where we replace 
$\phi^n$ by $\phi^{n+1/2}$.
\begin{eqnarray}
	\tilde{c} &=& c^n + \frac{\Delta t}{2}L_y\left(\phi_y^{n+1/2}\right)\tilde{c} + \frac{\Delta t}{2}L_x\left(\phi_x^{n+1/2}\right)c^n \\
	{c}^{n+1} &=& \tilde{c} + \frac{\Delta t}{2}L_y\left(\phi_y^{n+1/2}\right)\tilde{c} + \frac{\Delta t}{2}L_x\left(\phi_x^{n+1/2}\right)c^{n+1} 
\end{eqnarray}
where, as before, the solution to the Poisson equation is obtained with Fourier spectral method. 

This correction is very effective because it improves the order of accuracy with almost no additional cost than the first order scheme. 

\subsubsection{Accuracy test}
In this section we verify the expected accuracy of the three different versions of ADI we discusse before. The space discretization of the Eqs.~(\ref{eq_ADI}-\ref{eq_ADI_in}) is described in Section \ref{section_space_discrete} with number of points for each direction $N = 128$, and here we define the other quantities of the model:
\begin{eqnarray}
	c_{\rm A}(x,y) &=& \exp\left(-\frac{\sin(x/2)^2 + \sin(y/2)^2}{2\sigma^2}\right) \\ \nonumber
	c_{\rm in}(x,y) &=& \exp\left(-\frac{\sin((x-1.5)/2)^2 + \sin((y-1.5)/2)^2}{2\sigma^2}\right),
\end{eqnarray}
In our tests the diffusion coefficient is $\mu = 0.1$, and $\sigma = 0.1$. 
\begin{figure}
	\centering
	\begin{minipage}
		{.49\textwidth}
		\centering
		\includegraphics[width=\textwidth]{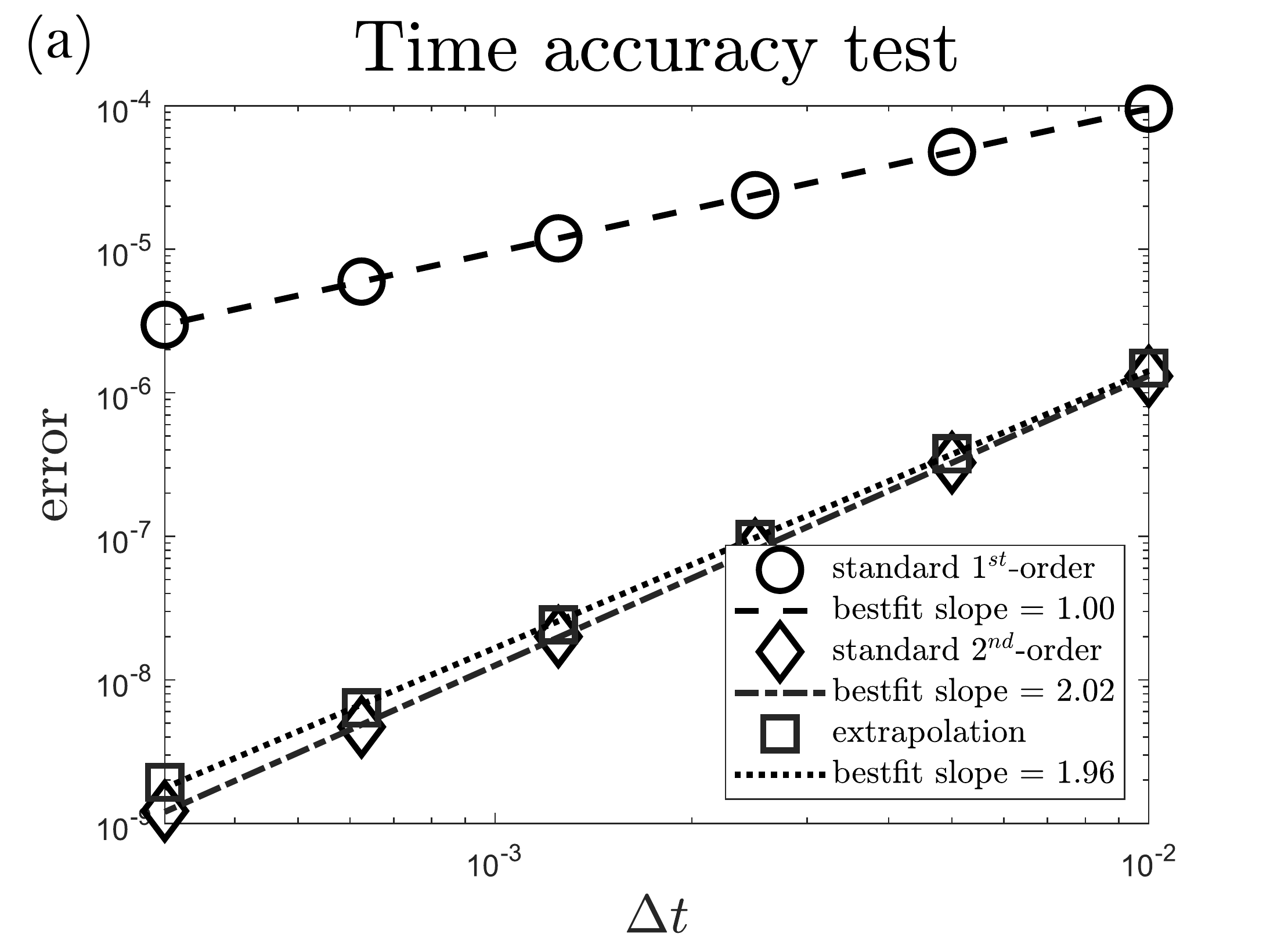}
	\end{minipage}
	\begin{minipage}
		{.49\textwidth}
		\centering
		\includegraphics[width=\textwidth]{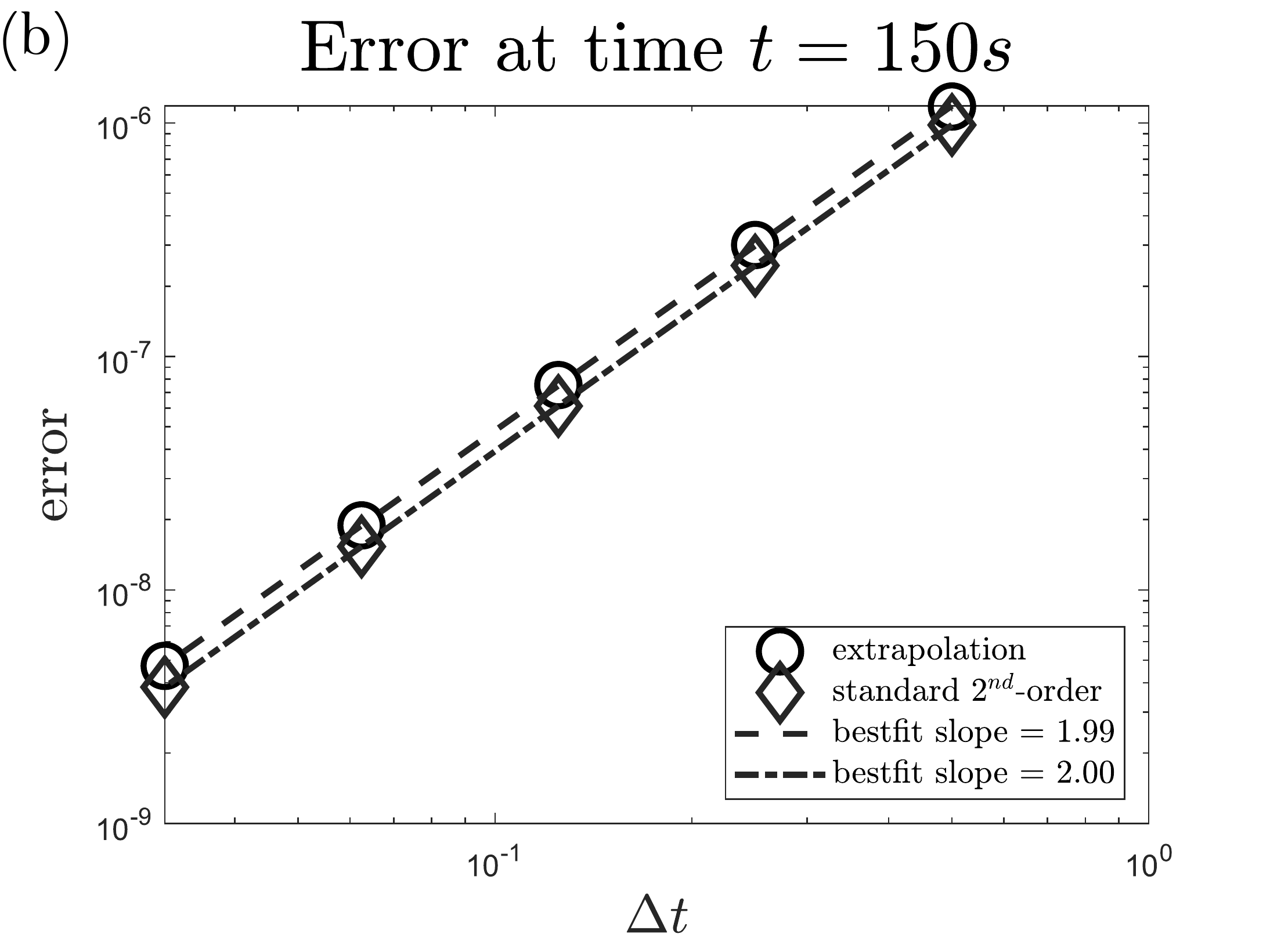}
	\end{minipage}
	\caption{\textit{{Time accuracy tests for the three different ADI methods we describe  {in this section}, at time $t=0.1$ with a reference solution (a) and at time $t=150$ by Richardson extrapolation (b), where we compare the two second-ordered methods. For the reference solution in (a) $\Delta t_{\rm ref} = 10^{-7}$. 
	}}}
	\label{figure_accuracy}
\end{figure}

In order to test the accuracy in time of the method, we compute a reference solution solving the numerical method described in Eqs.~(\ref{eq_standard_ADI}-\ref{eq_standard_ADI_3}), with $\Delta t_{\rm ref} = 10^{-7}$. Then we calculate the relative error between the reference solution and different solutions of the models we propose. The results are summarized in Fig.~\ref{figure_accuracy} (a). The method based on the extrapolation technique is second order accurate as the standard (second order) one.

Secondly, to test the accuracy for larger times, we compare the two second order methods at time $t=150s$, where the error is obtained by applying Richardson extrapolation, as illustrated in Section \ref{app:Richardson}  (see Fig.~\ref{figure_accuracy} (b)). The proposed ADI method based on extrapolation is the one we use in Section `\ref{section_results} for the 3D computation for the full two carrier model.

\subsection{A semi-implicit treatment for PNP system}
In this section we consider a different formulation of the PNP system which allows a much more stable time discretization, thus allowing 
efficient computation also for very small values of $\varepsilon$. Here we consider the most challenging case $\mu = 0$ in Eq.~\eqref{eq_ADI}. 
The effect of the implicit treatment of the diffusion term indeed helps stability. 

Eq.~\eqref{eq_ADI} becomes
\begin{eqnarray}\label{eq_phi_sub0}
	c_t &=& \nabla \cdot(c \nabla\phi) \\ \label{eq_phi_sub}
	- \Delta \phi &=& \frac{c - c_{\rm A}}{\varepsilon}  
\end{eqnarray} 
which can be rewritten in non-conservative form:
\begin{eqnarray}
	c_t = \nabla c \cdot \nabla \phi + c \Delta \phi 
	\label{eq:nc1}
\end{eqnarray} 
Now we substitute the term $\Delta \phi$ using Eq.~\eqref{eq_phi_sub}, obtaining
\begin{eqnarray}
\label{eq_stability}
	c_t = \nabla \phi \cdot \nabla c + c \frac{c_{\rm A} - c}{\varepsilon}.
	\label{eq:nc2}
\end{eqnarray} 


A very efficient tool for the numerical treatment of equations containing both stiff and non stiff terms is provided by IMEX schemes. 

A first order IMEX scheme for system~(\ref{eq_phi_sub0}-\ref{eq_phi_sub}) is given by
\begin{eqnarray}
\label{eq_imex_semiimpl}
	c^{n+1} &=& c^n + \Delta t\, D_1^{\rm up}\, c^{n+1}\nabla\phi^n + \Delta t c^n \frac{c_{\rm A}(t^{n+1}) - c^{n+1}}{\varepsilon} \\ \nonumber
	-\Delta \phi^n &=& \frac{c^n - c_{\rm A}(t^n)}{\varepsilon}
\end{eqnarray}
where $D_1^{\rm up}$ is the discrete operator for the space derivatives with \textit{upwind} direction.

In order to avoid a very large convection term $\textbf{u} = -\nabla \phi$, one could start with a well prepared initial condition $c(x,y,0) = c_{\rm A}(x,y,0)$, so that the initial condition is close to equilibrium. As $\varepsilon \to 0$, $c(x,y,t)$ remains closer and closer to $c_{\rm A}(x,y,t)$. 
Notice that if $\varepsilon$ is small and the initial condition is not well prepared then there will be a fast transient that will bring $c$ close to $c_{\rm A}$. During the transient a small time step has to be used.

\subsection{A conservative scheme in non conservative form}
Equations \eqref{eq:nc1} and \eqref{eq:nc2}  are written in non conservative form, however for smooth solutions they are equivalent to the corresponding equation in conservative form \eqref{eq_phi_sub}. 
It is possible to discretize \eqref{eq:nc1}  and \eqref{eq:nc2}  in space in such a way that conservation is guaranteed at semidiscrete level. 
Such conservative discretization takes the following form: 
\begin{equation}
    \frac{d{\bf c}}{dt} = D({\bf c},\Phi) + {\bf c} L \Phi
\end{equation}
where $\bf{c}$ and $\Phi$ denote, respectively, the concentration and the potential on the Cartesian grid $\Omega_h$, $L$ denotes the classical 5-point discrete Laplacian on a square grid, and the operator $D({\bf c},{  \Phi})$ is the following second order accurate discretization of the bilinear operator $\nabla f\cdot\nabla g$ applied to any two discrete functions $f,g$ defined on $\Omega_h$:
\begin{equation}
    D(f,g) = \frac{1}{2}(D_x^+ f  D_x^+ g +  D_x^- f  D_x^- g + D_y^+ f  D_y^+ g + D_y^-  f  D_y^- g )  
\end{equation}
with
\begin{align*}
    D_x^+ f_{ij} & \equiv \frac{f_{i+1,j}-f_{i,j}}{h}, \quad 
    D_x^- f_{ij} \equiv \frac{f_{i,j}-f_{i-1,j}}{h}, \\ 
    D_y^+ f_{ij} & \equiv \frac{f_{i,j+1}-f_{i,j}}{h}, \quad 
    D_y^- f_{ij} \equiv \frac{f_{i,j}-f_{i,j-1}}{h}. \quad 
\end{align*}
With the above definition, the semidiscrete scheme can be written as 
\begin{align}
    \frac{d{\bf c}}{dt} & = D({\bf c},\Phi) + {\bf c} 
    \frac{{\bf c}_A-{\bf c}}{\varepsilon}
    \label{eq:cnc1}\\
    L \Phi & = \frac{{\bf c}_A-{\bf c}}{\varepsilon}
    \label{eq:cnc2}
\end{align}
It may be shown that system \eqref{eq:cnc1}-\eqref{eq:cnc2} is conservative, i.e. 
\[
    \frac{d}{dt}\sum_{ij}c_{ij} = 0
\]
Furthermore, time discretizations of system \eqref{eq:cnc1}-\eqref{eq:cnc2} which are either fully explicit of fully implicit are exactly conservative as well.
IMEX schemes applied to system 
\eqref{eq:cnc1}-\eqref{eq:cnc2} are not exactly conservative, however, the conservation error depends only on time discretization, and is therefore smaller than the one obtained by a standard non-space conservative discretization (see Fig.~\ref{fig_cons_mass_2} (b)).

\subsubsection{Results}
In this section we show the improvements in stability  of the semi-implicit scheme defined in Eq.~\eqref{eq_imex_semiimpl} and the IMEX method applied to Eqs.(\ref{eq:cnc1}-\ref{eq:cnc2}). They are more efficient than the ADI method because of the restriction that the Debye length poses on the time step. With these schemes we are able to consider time steps few orders of magnitude larger than $\varepsilon$, also for negligible diffusion term. 

We define the  {discrepancy between} the background state and the numerical solution, $d_e(t)$, and the conservation mass error, $m_e(t)$: 
\begin{eqnarray}
\label{eq_discr}
    d_e(t) &= \frac{\sum_{ij} |c(t)-c_{\rm A}(t)|}{\sum_{ij} c_{\rm A}(t)}& \\
    \label{eq_cons_mass}
    m_e(t)&= \frac{\Big|\sum_{ij} (c(t)-c_{\rm A}(t))\Big|}{\sum_{ij} c_{\rm A}(t)}.& 
\end{eqnarray}

The expression for the background state we choose in our tests is
\[ c_{\rm A}(x,y,t) := \left(\cos(t)^2,\sin(t)^2\right)\cdot\left(
\begin{array}{c}
\exp\left(-(\sin((x-x_1)/2)^2+\sin((y-y_1)/2)^2)/(2\sigma^2)\right)\\
 \exp(-(\sin((x-x_2)/2)^2+\sin((y-y_2)/2)^2)/(2\sigma^2)) 
 \end{array}
 \right)
\]
with $x_1 = 1,x_2 = -1,y_1=y_2=0$. 

\begin{figure}[htb]
 	\centering
 		\begin{minipage}{.49\textwidth}
 		\centering
\includegraphics[width=\textwidth]{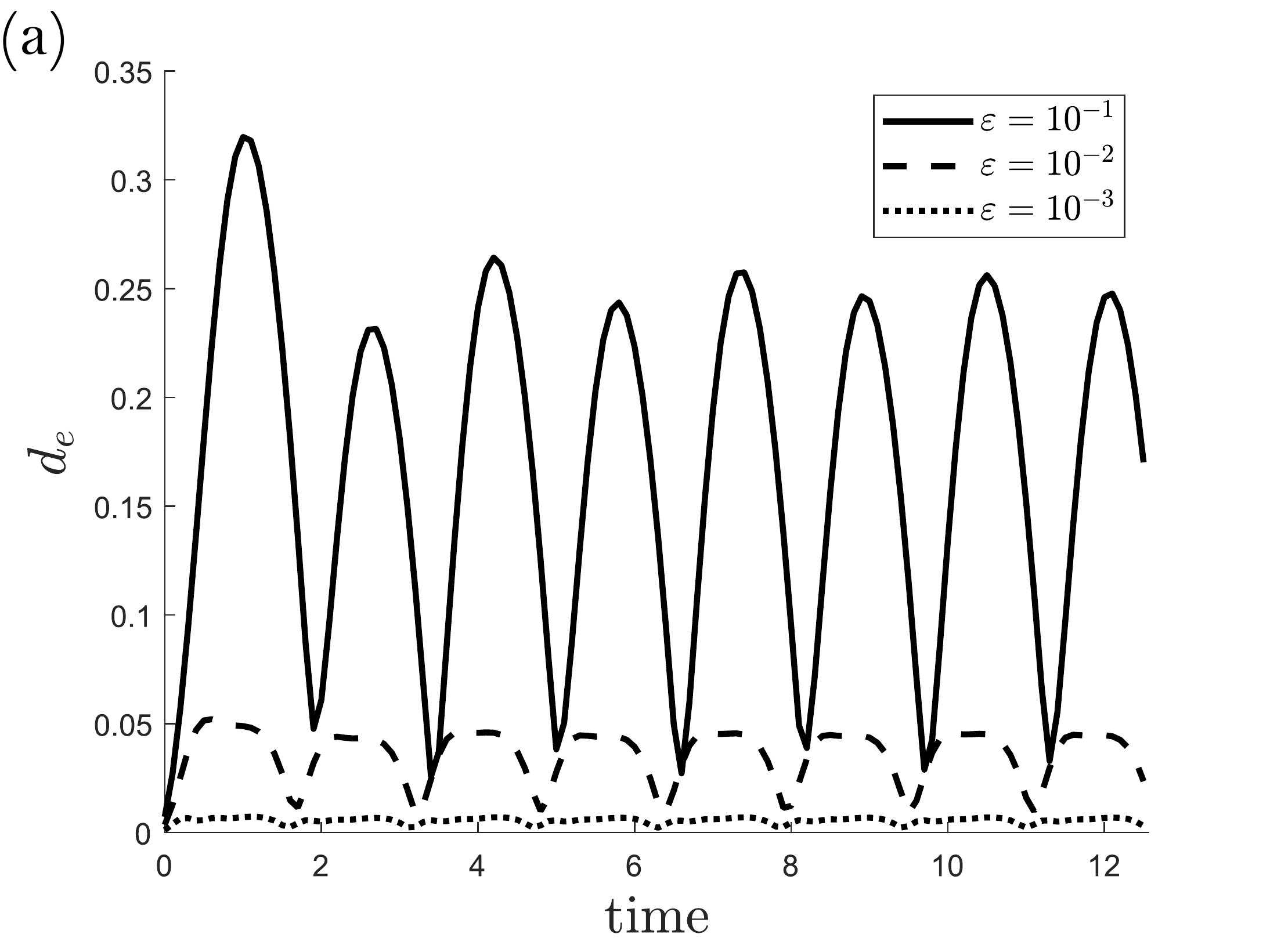}
	\end{minipage}
 	\begin{minipage}
 		{.49\textwidth}
 		\centering
 		\includegraphics[width=\textwidth]{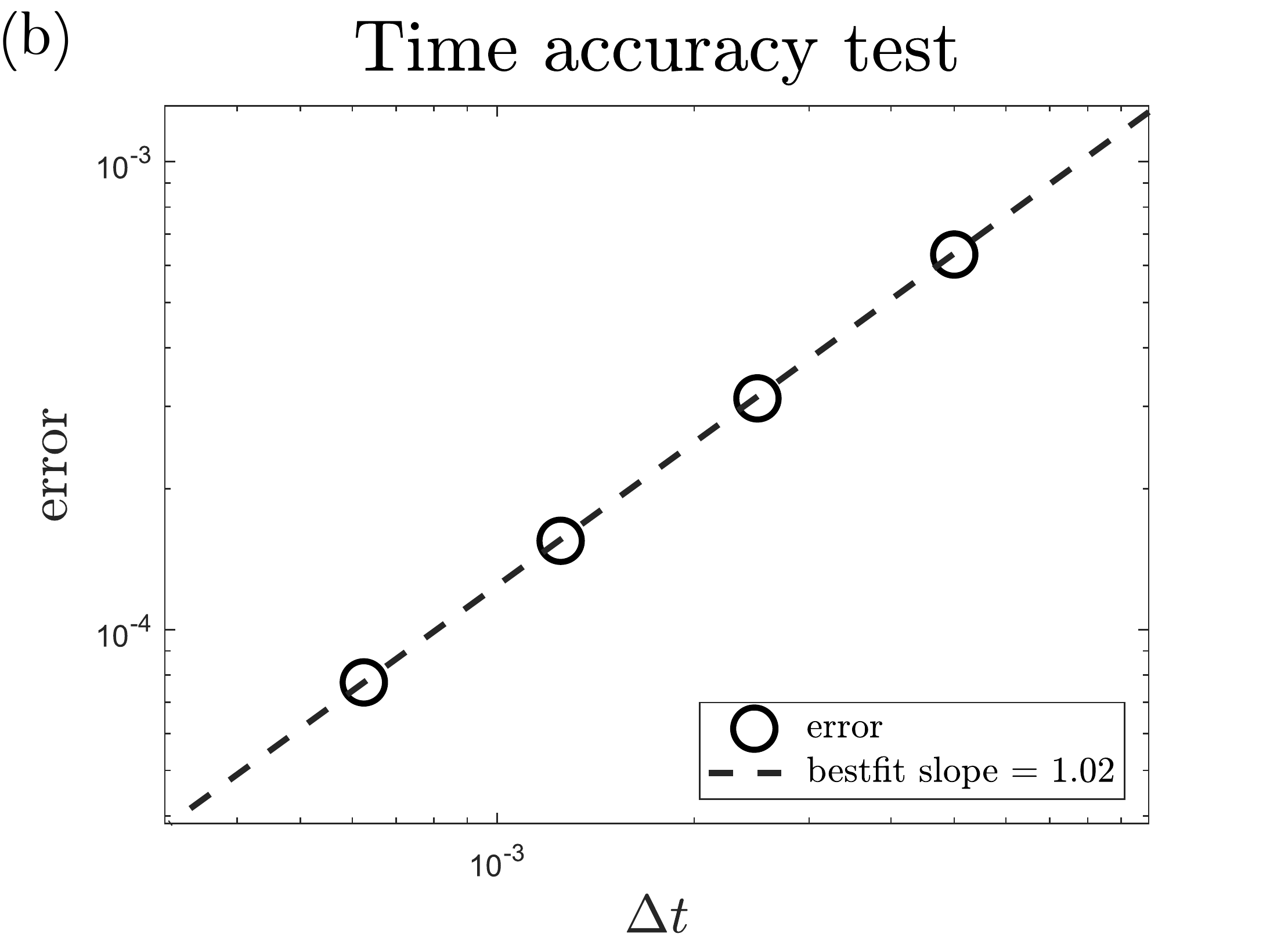}
 	\end{minipage}	\caption{\textit{ We show the evolution in time of the discrepancy, $d_e(t)$, defined in Eq.~\eqref{eq_discr} for different values of $\varepsilon \in \{10^{-1},10^{-2},10^{-3}$ \} and $\Delta t = 100\varepsilon$ (a) and the time accuracy test for the semi-implicit scheme (defined in Eqs.~\eqref{eq_imex_semiimpl}) (b) at time $t=0.1$. The reference solution is obtained with a time step $\Delta t_{\rm ref}  = 10^{-7}$. Grid resolution: $N=128$.}}
 	\label{fig_discr_mass}
\end{figure}

In Fig.~\ref{fig_discr_mass} (a) we show how the discrepancy, $d_e(t)$, strongly depends on $\varepsilon$ and how the concentration converges to the background as $\varepsilon \to 0$, while in panel (b) we show that the method defined in Eq.~\eqref{eq_imex_semiimpl} is first order in time. Here we plot (as before) the relative error between a reference solution obtained with $\Delta t_{\rm ref} = 10^{-7}$ and numerical solutions obtained with different time steps, at the final time $t=0.1$.

\begin{figure}[htb]
 	\centering
 		\begin{minipage}{.49\textwidth}
 		\centering
\includegraphics[width=\textwidth]{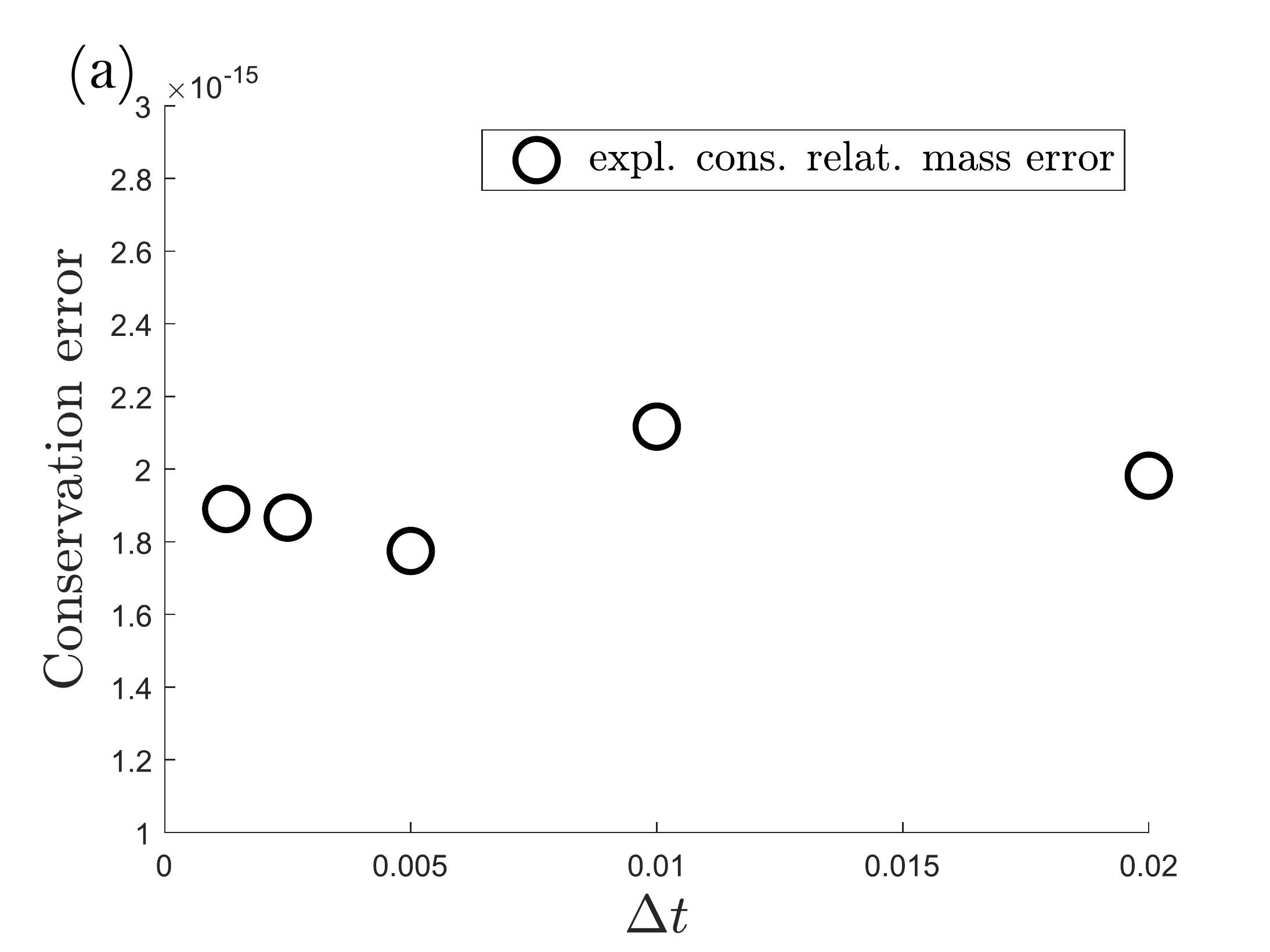}
	\end{minipage}
 	\begin{minipage}
 		{.49\textwidth}
 		\centering
	\includegraphics[width=\textwidth]{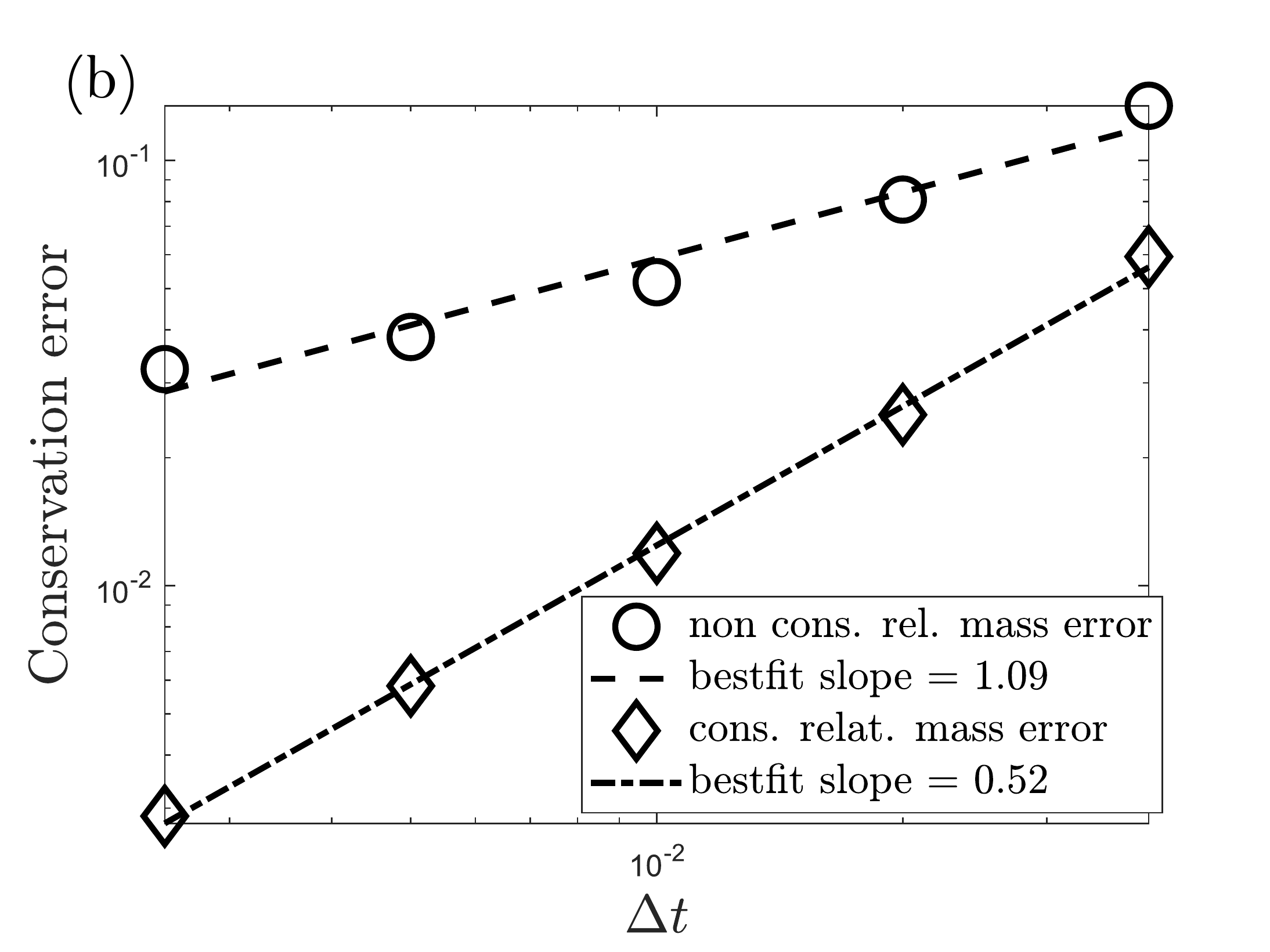}
 	\end{minipage}	
	\caption{\textit{ Here we show the average in time of the conservation error ${  \langle m_e \rangle } = \int_0^Tm_e(t)dt/T$, with $T = 2\pi$ for different values of $\Delta t$ for the explicit scheme (a) and the two semi-implicit schemes (b). For the explicit scheme the conservation of the total mass is guaranteed exactly, up to machine precision, while for the semi-implicit scheme written in conservative form the conservation error decreases with first order accuracy. Regarding the semi-implicit scheme in non conservative form, the conservation error does not decrease after a certain value of $\Delta t$, as we expected. Number of points in these tests is $N = 128$.}}
 	\label{fig_cons_mass_2}
\end{figure}

In Fig.~\ref{fig_cons_mass_2} we see the average in time of the conservation error, ${  \langle m_e \rangle }= \int_0^Tm_e(t)dt/T$, with $T = 2\pi$, versus $\Delta t$, for the scheme defined in Eqs.~\eqref{eq_imex_semiimpl} (circles panel (b)) and the explicit (panel (a)) and IMEX (diamonds panel (b)) schemes applied to Eq.(\ref{eq:cnc1}-\ref{eq:cnc2}). As we expected, the conservation of mass is exactly guaranteed for the explicit scheme (panel (a)), while for the semi-implicit schemes this does not happen (panel (b)). For the one written in non-conservative form, (see Eq.~\eqref{eq_imex_semiimpl}), the conservation error does not decrease further for sufficiently small values of $\Delta t$. We can also see that ${  \langle m_e \rangle }$ decreases with first order accuracy in $\Delta t$ for the semi-implicit scheme written in conservative form. 


In Table~\ref{table_semiimpl} we show how the discrepancy goes to zero with $\varepsilon$ and how the quantity depends on the fraction $\varepsilon/v_0$, where $v_0$ is the initial volume.

\begin{table}
	\centering
	\begin{tabular}{|c||c|c|c|c|} \hline
		\multicolumn{5}{|c|}{$v_0 = 10$} \\ \hline
		$\varepsilon$ & $10^{-2}$ & $10^{-3}$ & $10^{-4}$ & $10^{-5}$ \\ \hline
		error & 0.0096 & 5.05 $\cdot 10^{-4} $ & 3.52  $\cdot 10^{-5} $ & 8.07 $\cdot 10^{-6} $ \\ \hline \hline
	\multicolumn{5}{|c|}{$v_0 = 1$} \\ \hline
		$\varepsilon$ & $10^{-2}$ & $10^{-3}$ & $10^{-4}$ & $10^{-5}$ \\ \hline
		error &  0.183 & 0.0096 & 5.05  $\cdot 10^{-4} $ & 3.52 $\cdot 10^{-5} $ \\ \hline \hline
	\multicolumn{5}{|c|}{$v_0 = 0.1$} \\ \hline
		$\varepsilon$ & $10^{-2}$ & $10^{-3}$ & $10^{-4}$ & $10^{-5}$ \\ \hline
		error & - &  0.183 & 0.0096 & 5.08  $\cdot 10^{-4} $  \\ \hline
	\multicolumn{5}{|c|}{$v_0 = 0.01$} \\ \hline \hline
		$\varepsilon$ & $10^{-2}$ & $10^{-3}$ & $10^{-4}$ & $10^{-5}$ \\ \hline
		error & - & - &  0.183 & 0.0095 \\ \hline
	\end{tabular}
\caption{\textit{ Discrepancy $d_e$ on varying $\varepsilon $ and initial volume $v_0$ for the Eq.~\eqref{eq_imex_semiimpl}. From the values of the table we see the discrepancy strongly depends on the fraction $\varepsilon/v_0$. In these tests $\Delta t = 100\varepsilon$ and $N = 128$.}}
\label{table_semiimpl}
\end{table}

\section{Space discretization for the full PNP system}
\label{section_space_discretization}
In this section we describe the space discretization adopted in the numerical simulations of the PNP system. The scheme is second order accurate and we show the method to be conservative, therefore preserving, to machine precision, the total volume of both ion species (see Fig.~\ref{figure_cons_mass} (a)) and therefore their electric charges. In the same figure, panel (b), we also show the numerical solution is never negative, plotting the minimum of the concentration for different times, up to $t = 1000$.
\subsection{Space discretization}
\label{section_space_discrete}
\begin{figure}[htb]
	\centering
	\begin{minipage}
		{.49\textwidth}
		\centering
		\includegraphics[width=\textwidth]{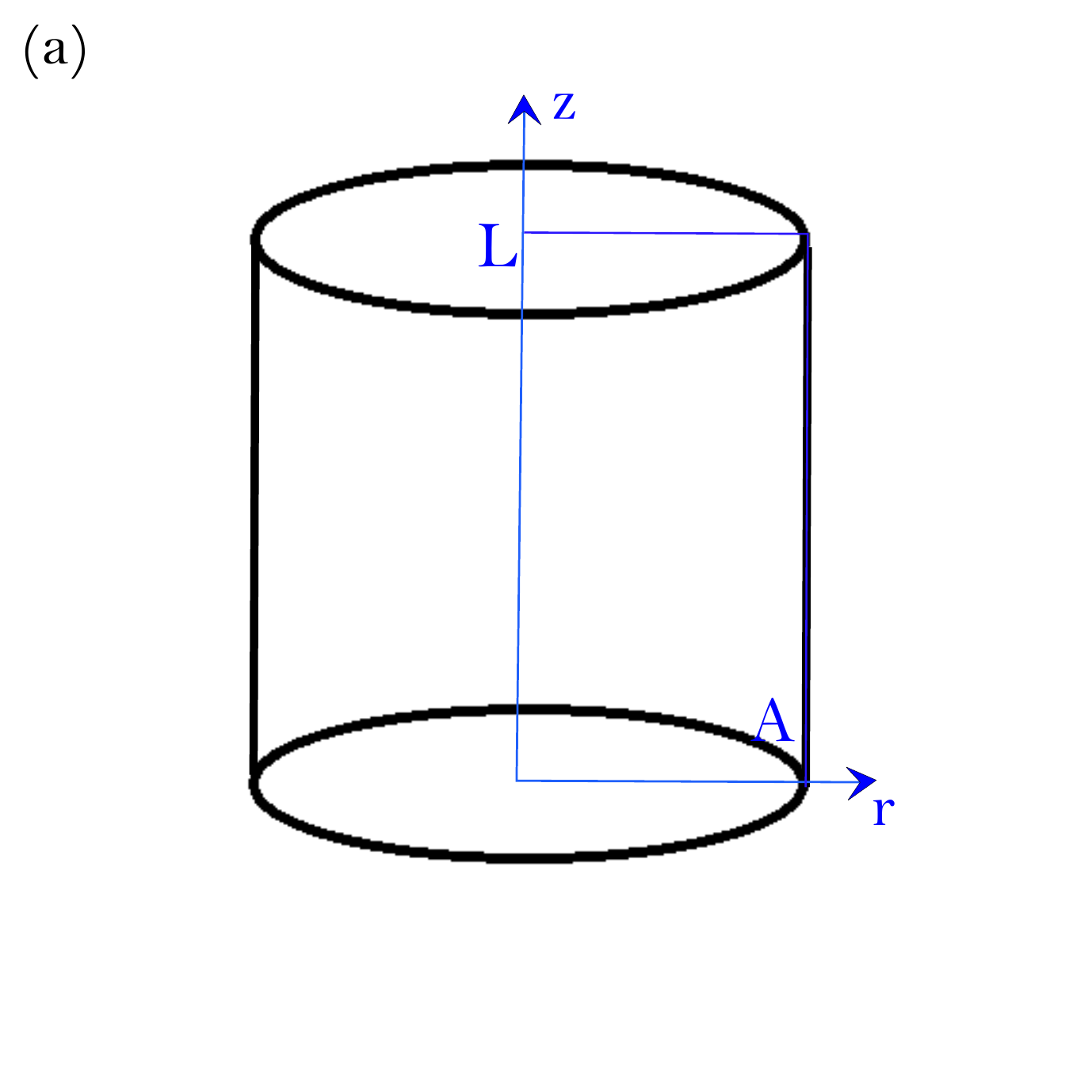}
	\end{minipage}
	\begin{minipage}
		{.49\textwidth}
		\centering
		\includegraphics[width=0.75\textwidth]{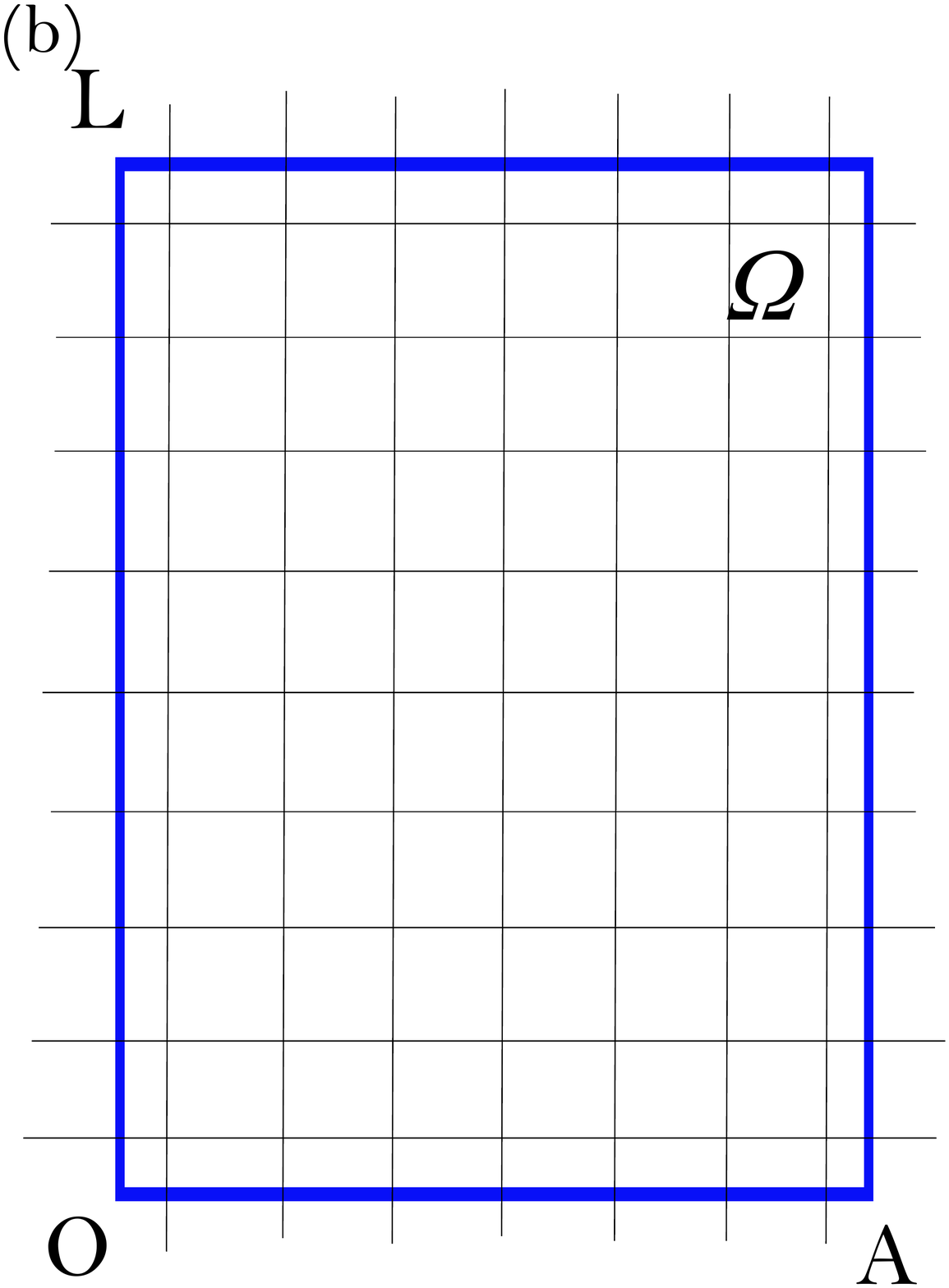}
	\end{minipage}
	\caption{\textit{Here we show the scheme of the domain in cylinder coordinates (a). The behaviour of the diffusion phenomenon is analogous to the case of a cube domain. For this reason we consider a cylinder instead. The rectangle of size $A\times L$ is the domain we are considering and on panel (b) we see the zoom-in with a cell-centered discretization in space. $A$ is the radius of the cylinder and $L$ its height. $O$ is the origin of the domain.}}
	\label{figure_domain}
\end{figure}

The original domain is a  {circular cylinder of height $L$ and radius $A$ (see Fig.~\ref{figure_domain} (a))}. The whole problem is therefore solved in three space dimension, assuming perfect cylindrical symmetry, both of the device and of initial and boundary conditions. 

The equations are written in cylindrical coordinates, and, taking advantage of \textit{axisymmetry}, the computational domain $\Omega$ is a two dimensional domain parameterized by $(r,z)$ coordinates, which span the rectangle $[0,A]\times[0,L]$ as we show in Fig.~\ref{figure_domain} (b). The computational domain $\Omega$ is then discretized by a uniform Cartesian mesh with spatial step $h := \Delta r = \Delta z$. We call $\Omega_h$ the discrete computational domain. The concentrations ${  c^\pm_{ij}\approx c^\pm(r_i,z_j)}$ and ${  C_{ij}\approx C(r_i,z_j)}$ are defined at the center of the cell $(i,j)$, therefore we have $r_i=(i-1/2)h, \, z_j=(j-1/2)h, \,(i,j)\in\{1,\dots,N_r\}\times\{1,\dots,N_z\}$, $h N_z = L$,  $h N_r = A$. 

\begin{figure}[htb]
	\centering
	\begin{minipage}
 		{.49\textwidth}
 		\centering
 		\includegraphics[width=\textwidth]{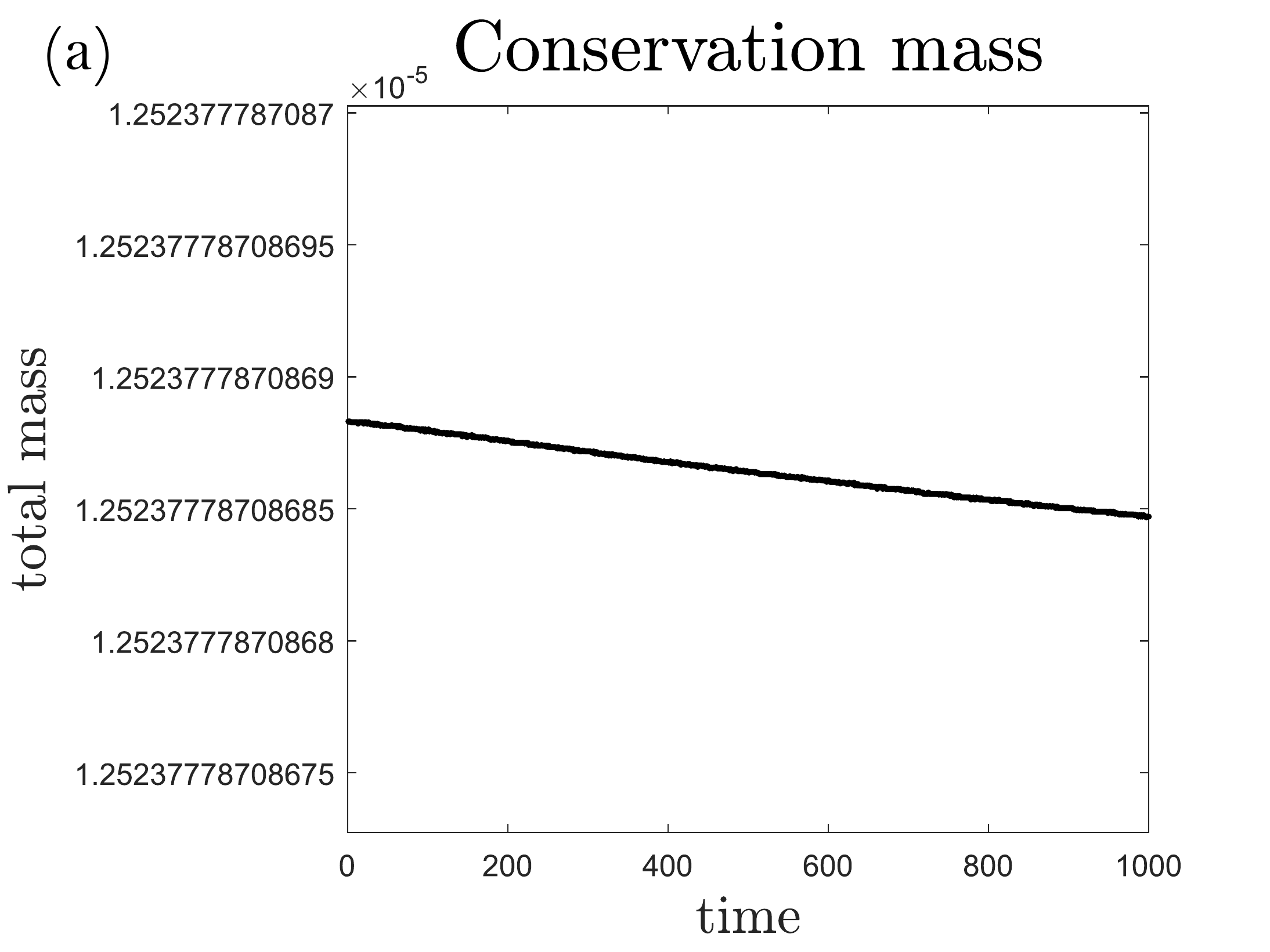}
 	\end{minipage}
 	\begin{minipage}
 		{.49\textwidth}
 		\centering
 		\includegraphics[width=\textwidth]{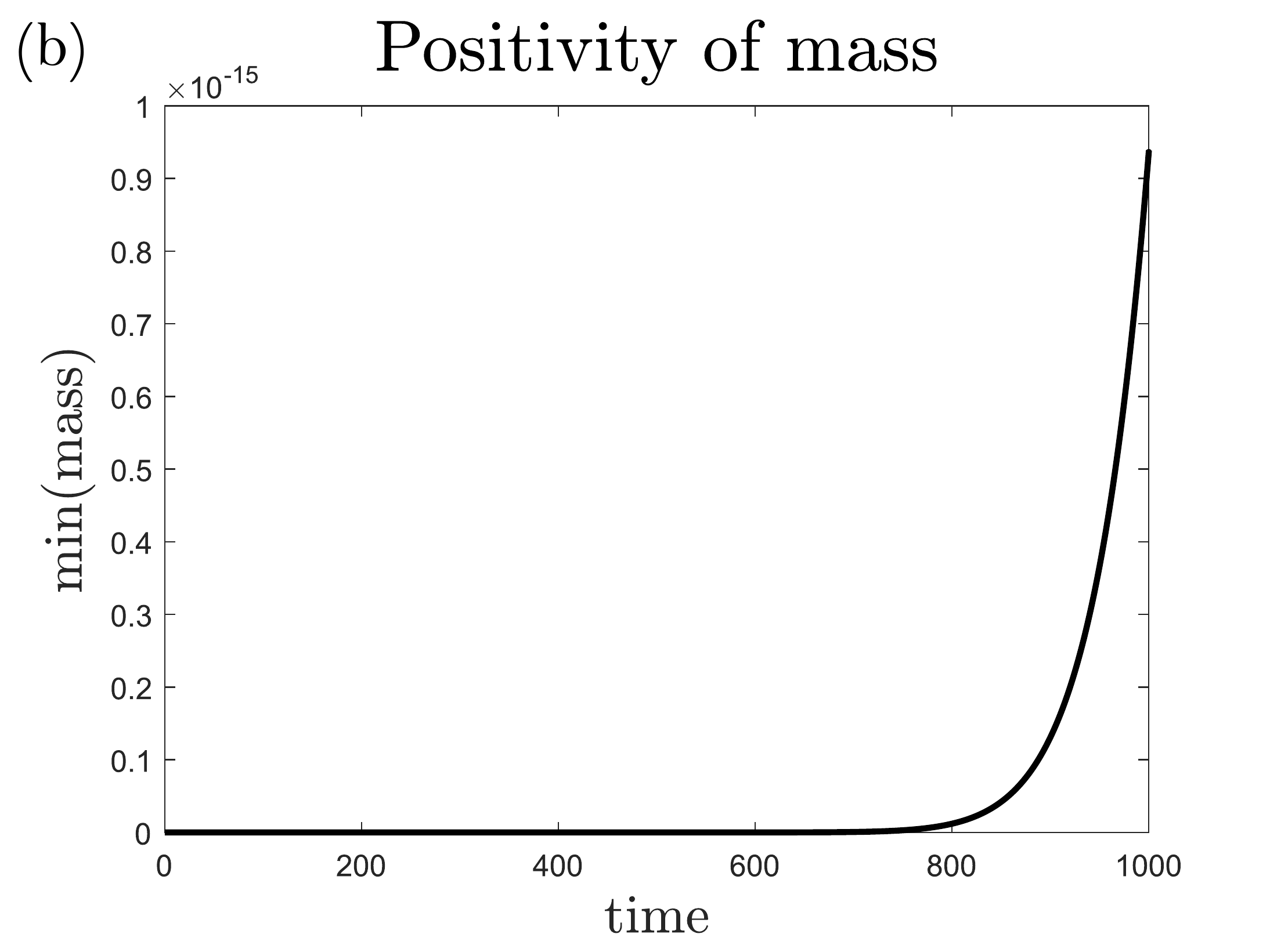}
 	\end{minipage}	\caption{\textit{Here we see the exact conservation of mass in time with machine precision (a) and the minimum of the solution versus time (b) to show the concentration is never negative. Thus the positivity of the solution is always guaranteed. In these plots $\varepsilon = 10^{-4}, h = 0.0133$ and $\Delta t  = 1{ .}$}}
	\label{figure_cons_mass}
\end{figure}

In order to obtain second order accuracy in space, we use central difference for the computation of the  {space derivatives}. 
Discretizing in space Eqs.~(\ref{equation_flux}-\ref{equation_flux_explicit}) we have:
	\begin{eqnarray}
		\nonumber
		\displaystyle \frac{\partial  c_{ij}^\pm}{\partial t}&=& -\frac{1}{r_{i}}\frac{r_{i+1/2}\,J_{i+1/2\,j}^{\pm,r}-r_{i-1/2}\,J_{i-1/2\,j}^{\pm,r}}{\Delta r} - \frac{J_{i\,j+1/2}^{\pm,z}-J_{i\,j-1/2}^{\pm,z}}{\Delta z},\\ \nonumber
		\displaystyle J_{i+1/2\,j}^{\pm,r}&=&-D\left(\frac{ c_{i+1\,j}^\pm- c_{ij}^\pm}{\Delta r} +  c^{ \pm}_{i+1/2\,j}\frac{U_{i+1\,j}-U_{i\,j}}{\Delta r}\right) \quad \forall i\neq1,N,\, \forall j \\\nonumber
		\displaystyle &=& -D\left(\frac{ c_{i+1\,j}^\pm- c_{ij}^\pm}{\Delta r} + \frac{ c_{ij}^\pm+ c^{ \pm}_{i+1\,j}}{2}\frac{U_{i+1\,j}-U_{ij}}{\Delta r}\right)  \quad \forall i\neq 1,N,\, \forall j\\
		\label{cons_mass_cont}
		\displaystyle J^{\pm,r}(r_{1/2},z_j) &=& 0,\quad \forall j: \quad J^{\pm,r}(r_i,z_{1/2}) = 0,\quad \forall i\\ 	\label{cons_mass_cont_2}
		\displaystyle J^{\pm,z}(r_{N+1/2},z_j) &=& 0, \quad \forall j; \quad J^{\pm,r}(r_i,z_{N+1/2}) = 0,\quad \forall i
	\end{eqnarray}
	where the discretization of $J^{\pm,z}$ is  {omitted} because it is analogue to the  {one of the} $r$ component.
We choose a cell centered discretization because it is easier to implement the boundary condition  {and} to guarantee the exact conservation of mass (see Fig.~\ref{figure_cons_mass} (a))  {which derives} from the zero boundary condition for the flux and here we prove it as follows:
	\begin{eqnarray}
		\nonumber
		\displaystyle \sum_{i,j}\frac{d c_{ij}^\pm}{dt} &=& -\sum_{j=1}^{N_z}\sum_{i=1}^{N_r}\frac{r_{i+1/2}J^{\pm,r}_{i+1/2\,j}-r_{i-1/2}J^{\pm,r}_{i-1/2\,j}}{r_{i}\Delta r }\,2\,\pi\, r_{i}\,\Delta r  \Delta z \\ && - \sum_{i=1}^{N_r}\sum_{j=1}^{N_z} \frac{J^{\pm,z}_{i\,j+1/2}-J^{\pm,z}_{i\,j-1/2}}{\Delta z}\,2\,\pi\, r_{i}\,\Delta z \Delta r\\ \nonumber
		\displaystyle &=& -\sum_{j=1}^{N_z}\left(A\cdot J^{\pm,r}(A,z_j)-0\cdot J^{\pm,r}(0,z_j)\right) \Delta z  \\ &&
		-\sum_{i=1}^{N_r}\left(J^{\pm,z}(r_i,L)-J^{\pm,z}(r_i,0)\right) \,2\,\pi\, r_i \Delta r
	\end{eqnarray}
	the right side of the last equation is equal to zero because of the conditions defined in Eqs.~(\ref{cons_mass_cont}-\ref{cons_mass_cont_2}).
The diffusion term suggests a central differencing scheme, which is second order accurate, and it is 
stable even in presence of a drift term, provided the so called \textit{mesh P\'{e}clet number} is sufficiently small \cite{Wesseling2023600}. 

We therefore choose a space step $h$ such that 
\begin{equation}
	\left|{\partial_r U^\pm}\right| < 2/h, \quad \left|{\partial_z U^\pm}\right| < {2}/{h}.
\end{equation}

Notice that if this condition is not satisfied, this means that the potential is not well resolved, therefore the accuracy of the whole procedure becomes questionable. Realistic potentials have a much shorter range than the one considered in this paper. In order to resolve the space scales one should use a much finer mesh, possibly adopting Adaptive Mesh Refinement techniques (AMR) \cite{hao2020adaptive}. An alternative would be to construct a multiscale model, which describes the effect of the potential through a suitable boundary condition. This approach is currently under investigation \cite{multiscale_mod}.

\section{Results and discussion}
\label{section_results}
In this section we perform several simulations with the aim of studying the effect of the various parameters. In particular, we check the agreement of PNP model with QNL model as the Debye length becomes smaller and smaller, and compare simulations with experiments. Initial conditions are defined in Eqs.~\eqref{equation_IC},~\eqref{QNL-IC} and adopted parameters are reported in Table~\ref{table_parameters}. 
\begin{table}
	\centering  
	\begin{tabular}{|c|c|c|c|c|c|}
		\hline
		Symbol & value & Symbol & value & Symbol & value\\ 
		\hline\hline 
		$a_1$ & 4 & $a_2$ & 6 & $a_3$ & 3\\ \hline
		$b_1$ & $30\,mm^{-2}$ & $b_2$ & $30\,mm^{-2}$ & $b_3$ & $10\,mm^{-2}$ \\ \hline		
		$k_BT$ & $4.14 \times 10^{-21}J$ &$A$ & $4\,mm$ & $L$ & $ 6\, mm$  \\ 
		\hline 
		$D_0$ & $10^{-9}m^2 s^{-1}$ & $D^+/D_0$ & $1.5$  & $D^-/D_0$ & $0.5$\\
		\hline 
		$\epsilon_0$ & $8.8541\times10^{-12}\,F m^{-1}$ & $\epsilon_r$ & $78$ & $\sigma$ & $0.4$\\
		\hline		
		$m_0$ & $10^{-3}\,$ Kg$\, mol^{-1}$ & $m^+$ & $23$ & $m^-$ & $265$\\
		\hline
		$q$ & $1,602\times 10^{-19}C$  & $z_c$ & $2.6\,mm$ &  $H$ & $4\,mm$ \\
		\hline
		$N_A$ & $6.022\times10^{23}mol^{-1}$ & $v_0^+$ & $ 10^{-6} $ &$\rho$ & $10^{3}\,Kg\, m^{-3}$\\
		\hline 
	\end{tabular}
	\caption{\textit{Parameters involved.}}
	\label{table_parameters}
\end{table}
\begin{figure}[htb]
	\centering	
	\begin{minipage}
		{.48\textwidth}
		\centering
		\includegraphics[width=\textwidth]{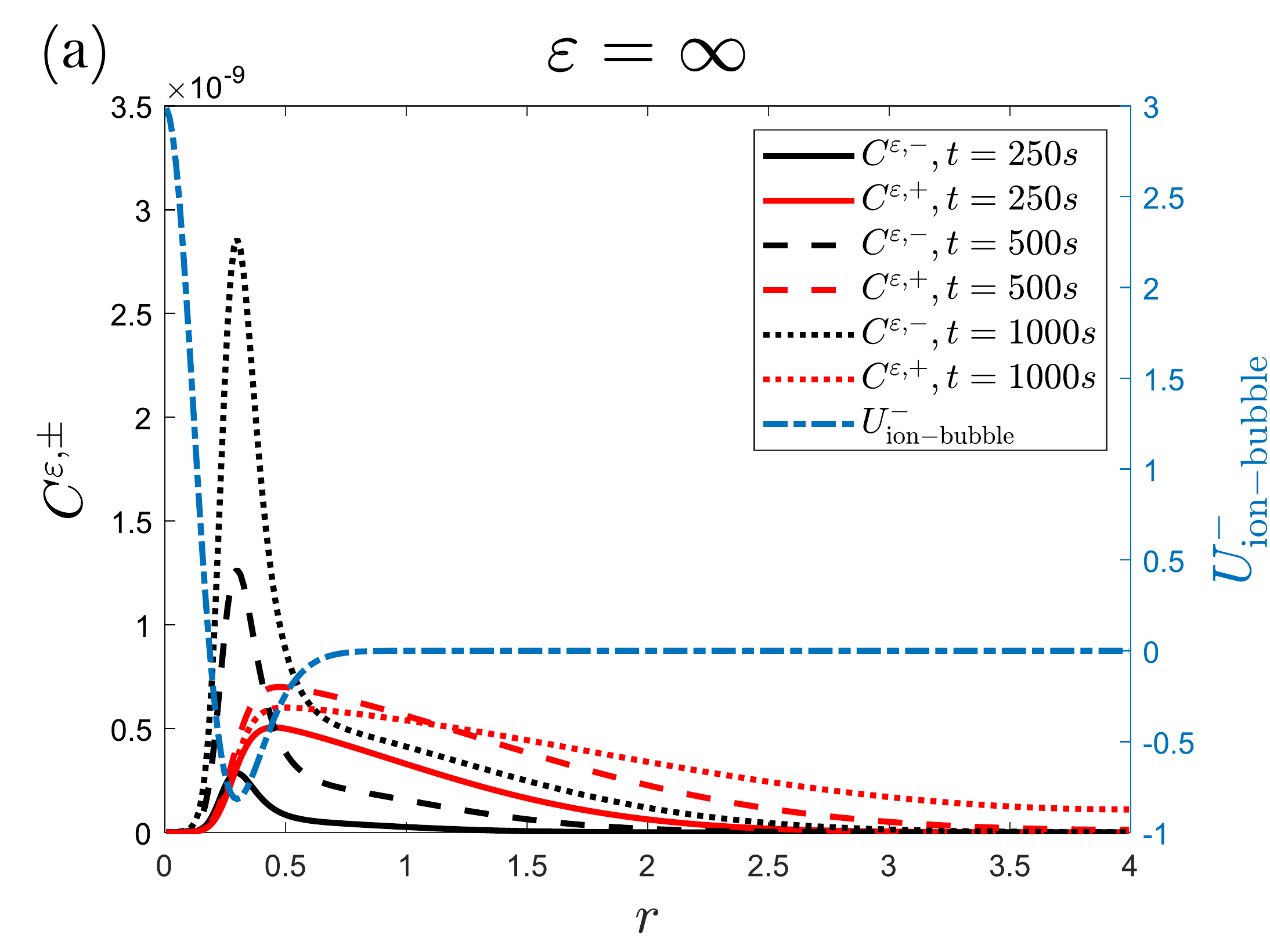}
	\end{minipage}
	\begin{minipage}
		{.48\textwidth}
		\centering
		\includegraphics[width=\textwidth]{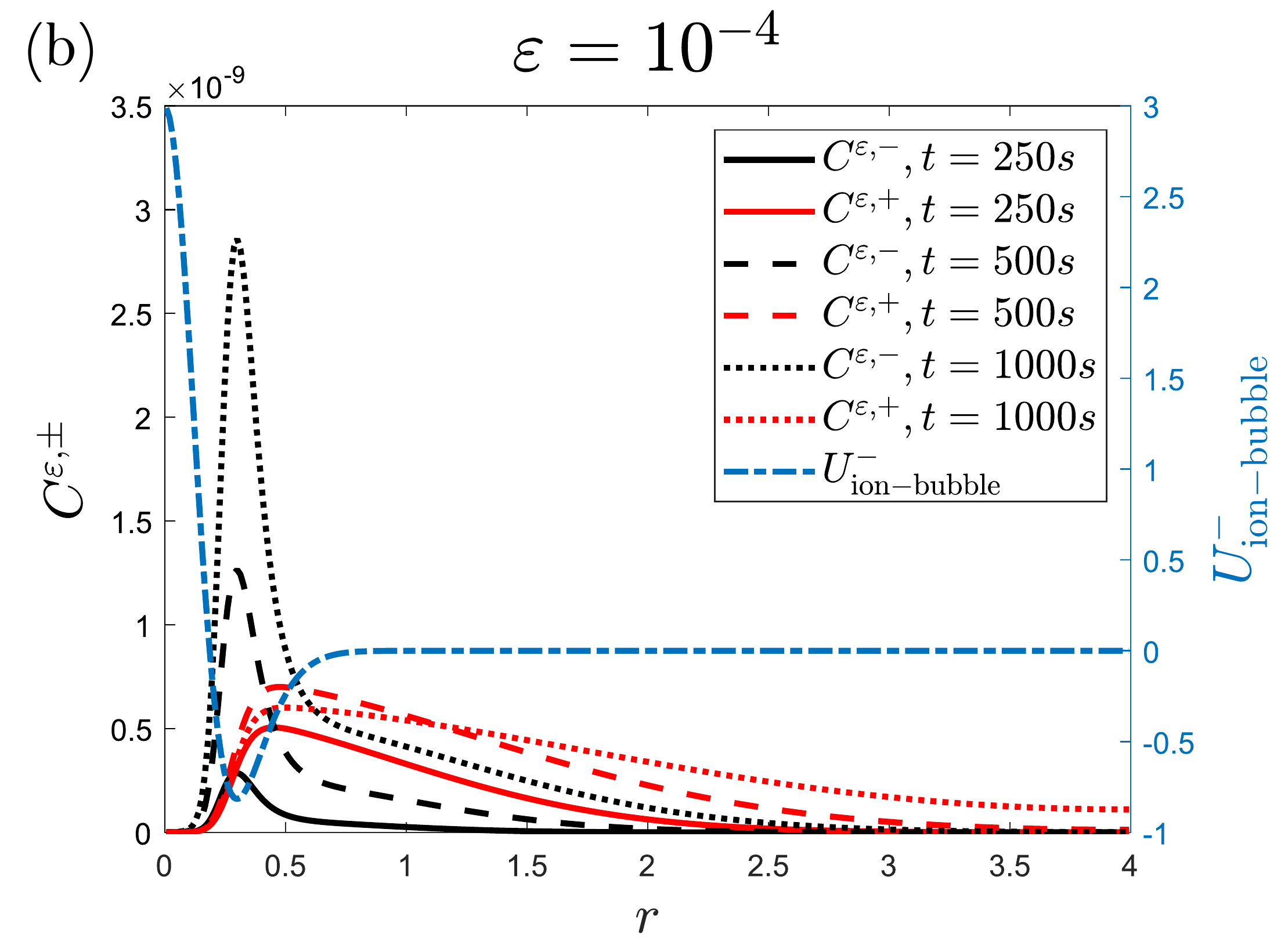}
	\end{minipage}
	\begin{minipage}
		{.48\textwidth}
		\centering
		\includegraphics[width=\textwidth]{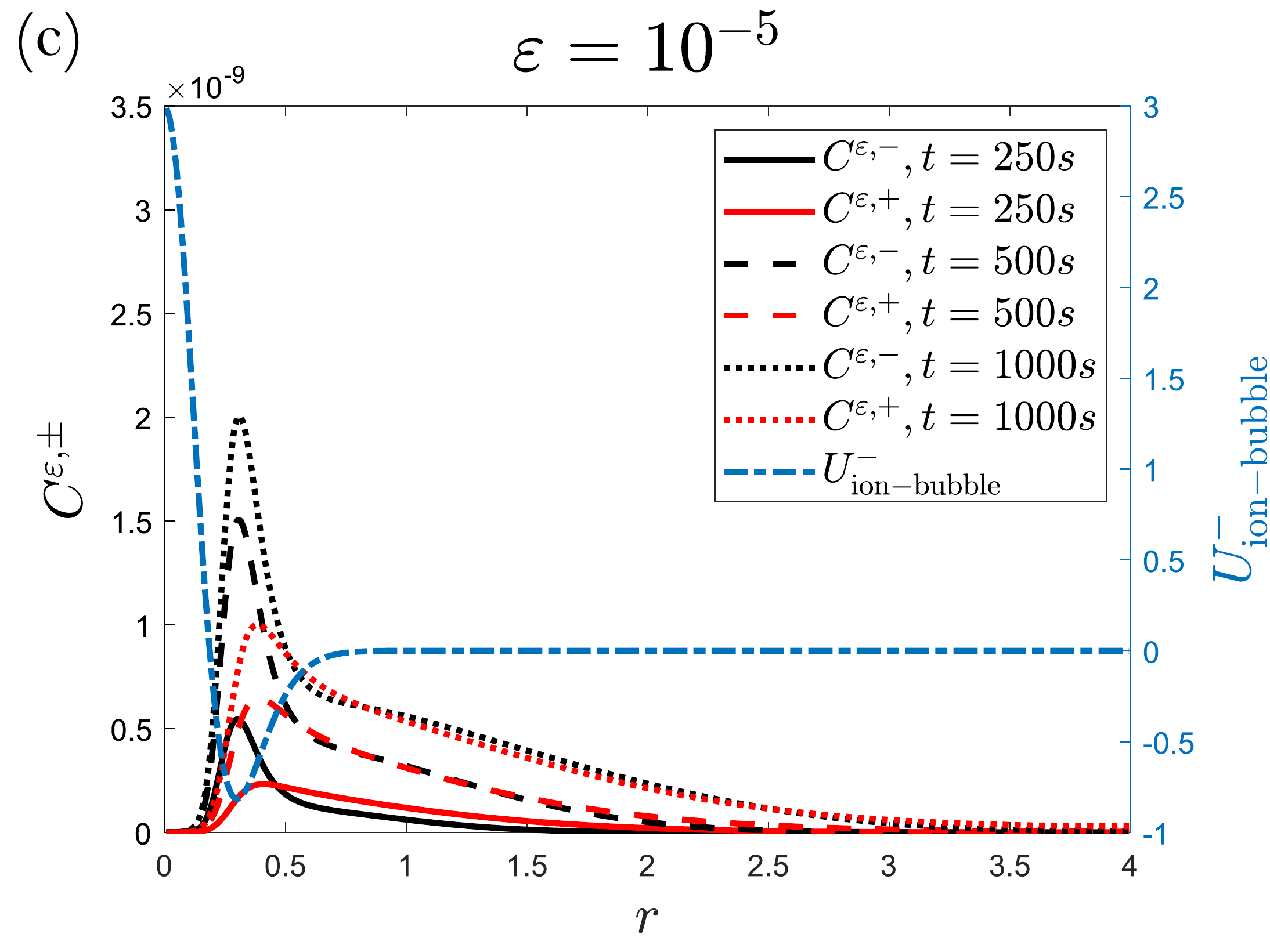}
	\end{minipage}
	\begin{minipage}
		{.48\textwidth}
		\centering
		\includegraphics[width=\textwidth]{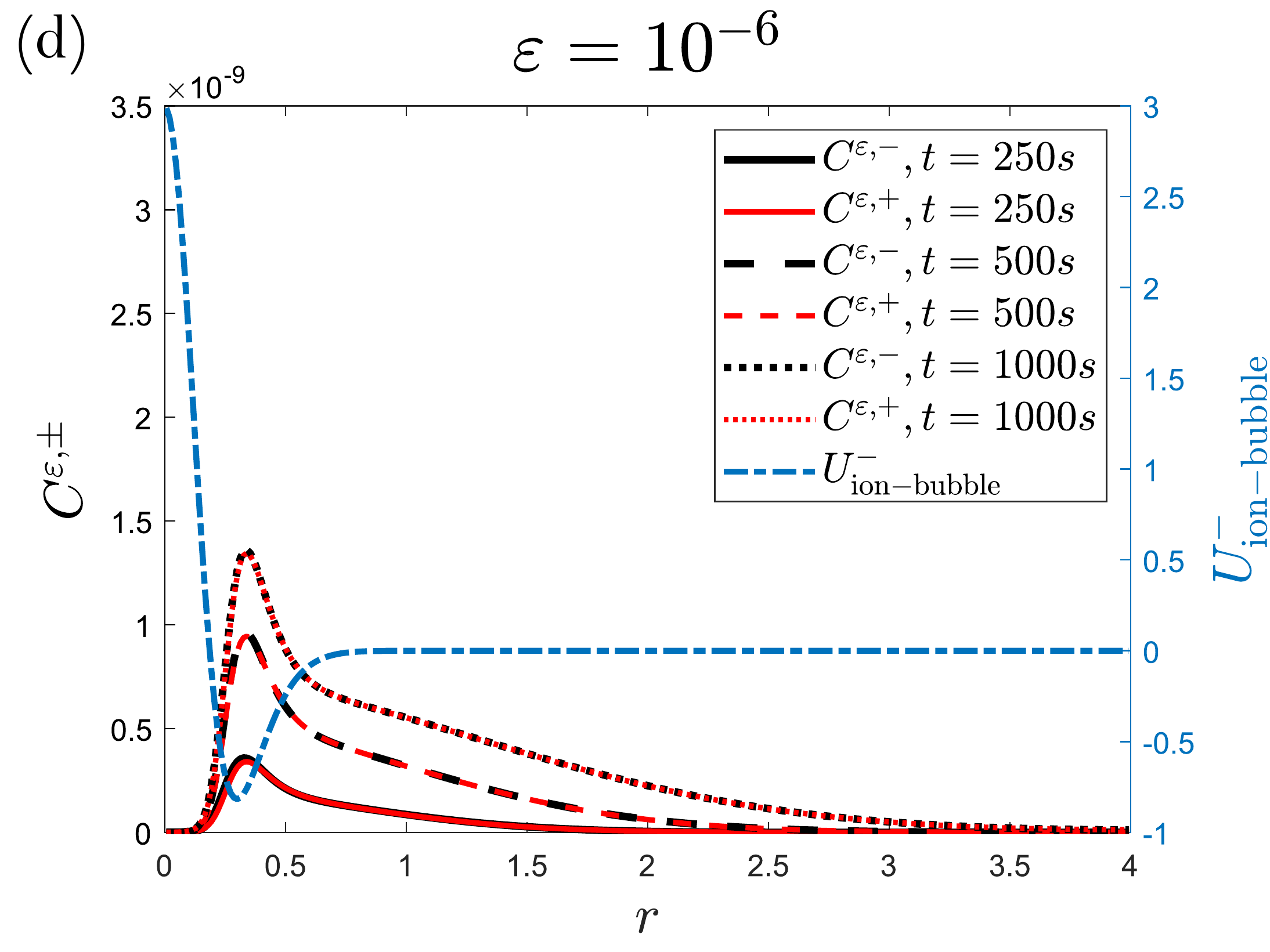}
	\end{minipage}
	\caption{\textit{Radial distribution of the ion charges at $z = z_c$, $C^{\varepsilon,\pm} = c^{\varepsilon,\pm}/m^\pm$ (black-red, left scale), solutions of PNP model for different times 
			and for various values of $\varepsilon$. 
			The  {dot-}dashed lines represent the potential for the anions (right scale), as we see in Fig.\ref{QNL_ext_plot} left panel. 
			Grid resolution: $h = 0.0133$ and time step $\Delta t = 1$ for (a),(b), $\Delta t = 0.1$ for (c) and $\Delta t = 0.005$ for (d).}}
	\label{1Dplots}
\end{figure}

Physical value of $\varepsilon \approx 10^{-8}$  {in Eq.~\eqref{equation_lap_Ueps} } is prohibitively low, and would make the system too stiff for numerical treatment. For this reason we performed the calculation using values of $\varepsilon$ which are larger than realistic ones, and study how the solution depends on $\varepsilon$.

In Fig.~\ref{1Dplots} we show the profile of the ion charge density at $z = z_c$ as a function of $r$, for different values of time $t\in\{250s,\,  500s,\,  1000s\}$ and of $\varepsilon \in \{\infty\, (a),10^{-4}\, (b),10^{-5}\, (c), 10^{-6}\, (d)\}$. By $\varepsilon = \infty$ we mean to switch off the electrostatic term. The ion charge density ($C^{\varepsilon,\pm}:=c^{\varepsilon,\pm}/m^\pm$) increases near the bubble because of the strong attraction of the potential $U^-_{\rm{ion-bubble}}$ that is represented by the  {dot-}dashed line in the same figure, right scale. Note that choosing $\varepsilon = 10^{-4}$ the effect of the electrostatic potential is negligible: the ions diffuse almost independently of each other, with the cations diffusing faster. Choosing $\varepsilon = 10^{-5}$ the effect of the electrostatic term is stronger and the two profiles start to get closer while for $\varepsilon = 10^{-6}$ they are almost overlapped (panel (d)), justifying the use of quasi-neutrality for the more realistic value $\varepsilon = 10^{-8}$. 
\begin{figure}[htb]
	\centering
	\begin{minipage}
		{.48\textwidth}
		\centering
	\includegraphics[width=1\textwidth]{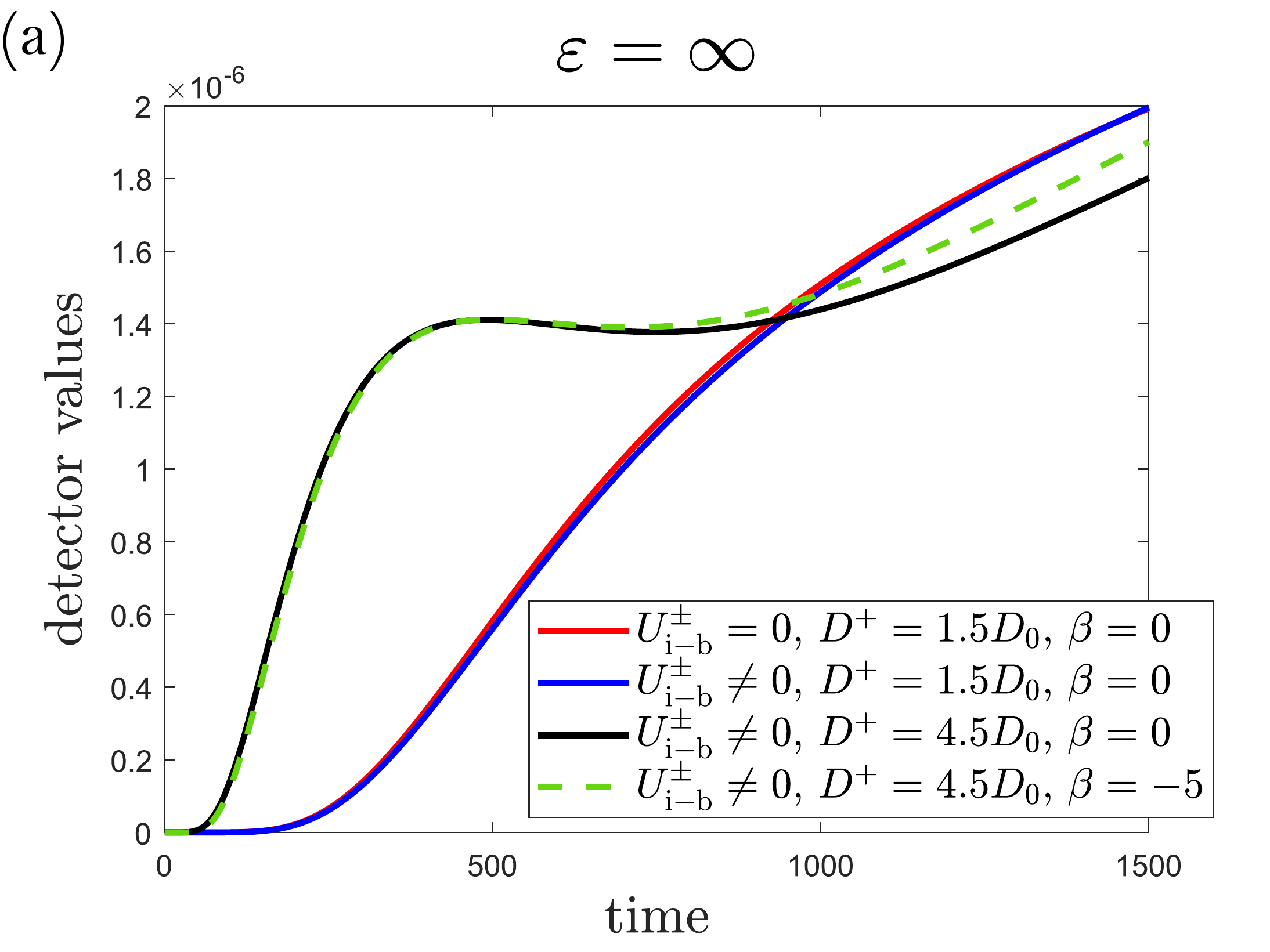}
	\end{minipage}
	\begin{minipage}
		{.48\textwidth}
		\centering
	\includegraphics[width=1\textwidth]{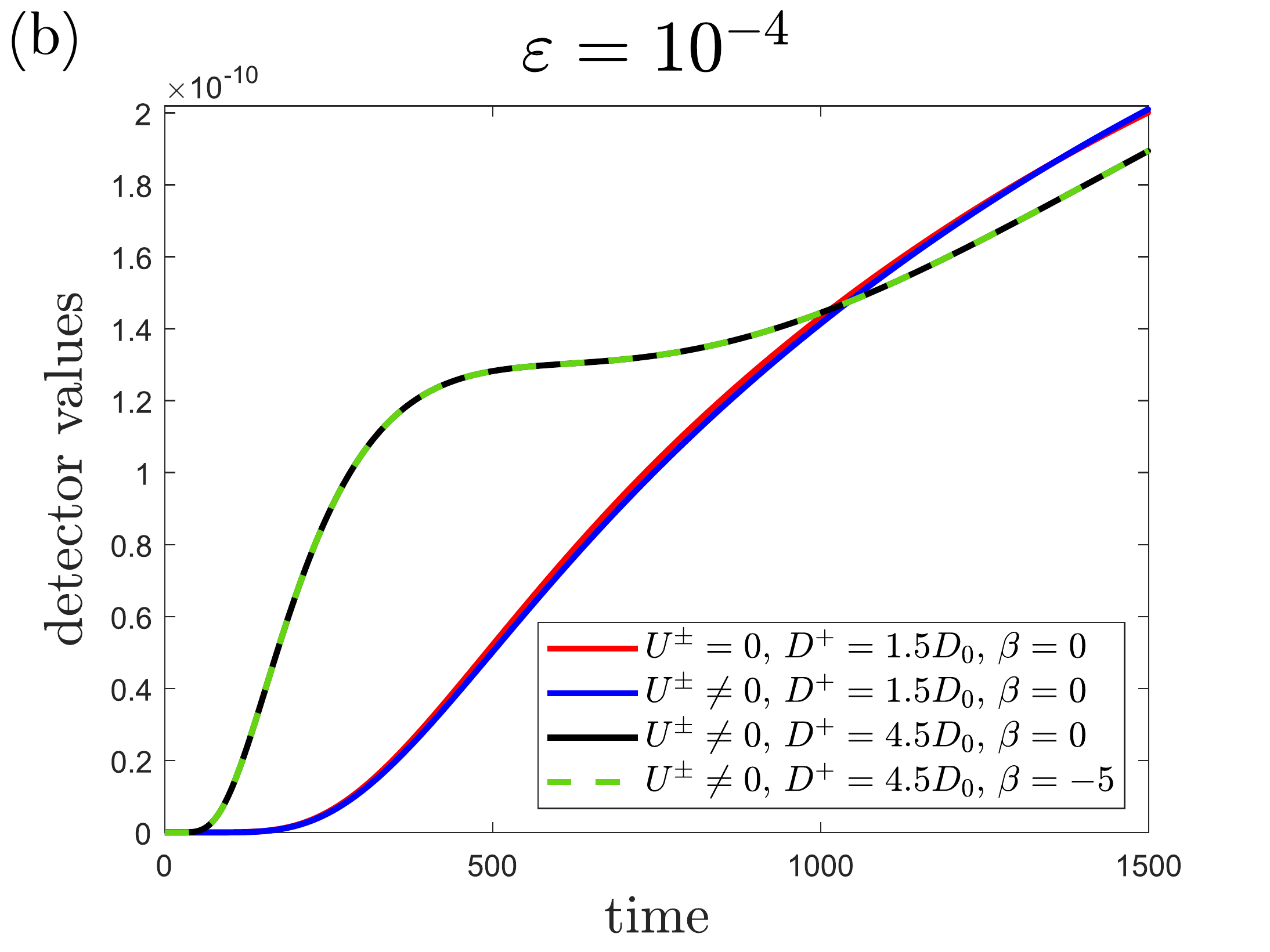}
	\end{minipage}
	\begin{minipage}
		{.48\textwidth}
		\centering
		\includegraphics[width=1\textwidth]{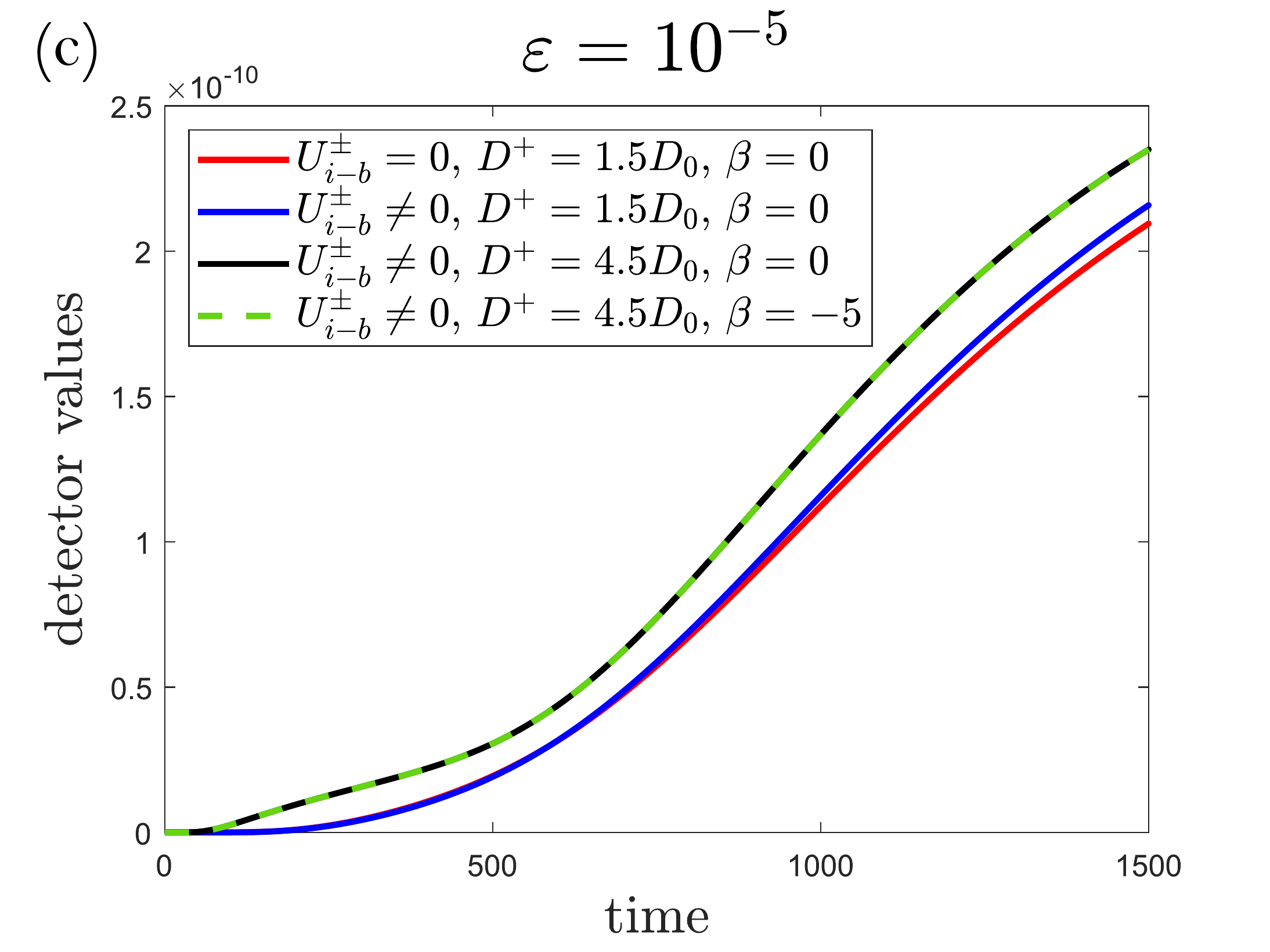}
	\end{minipage}
	\begin{minipage}
		{.48\textwidth}
		\centering
		\includegraphics[width=1\textwidth]{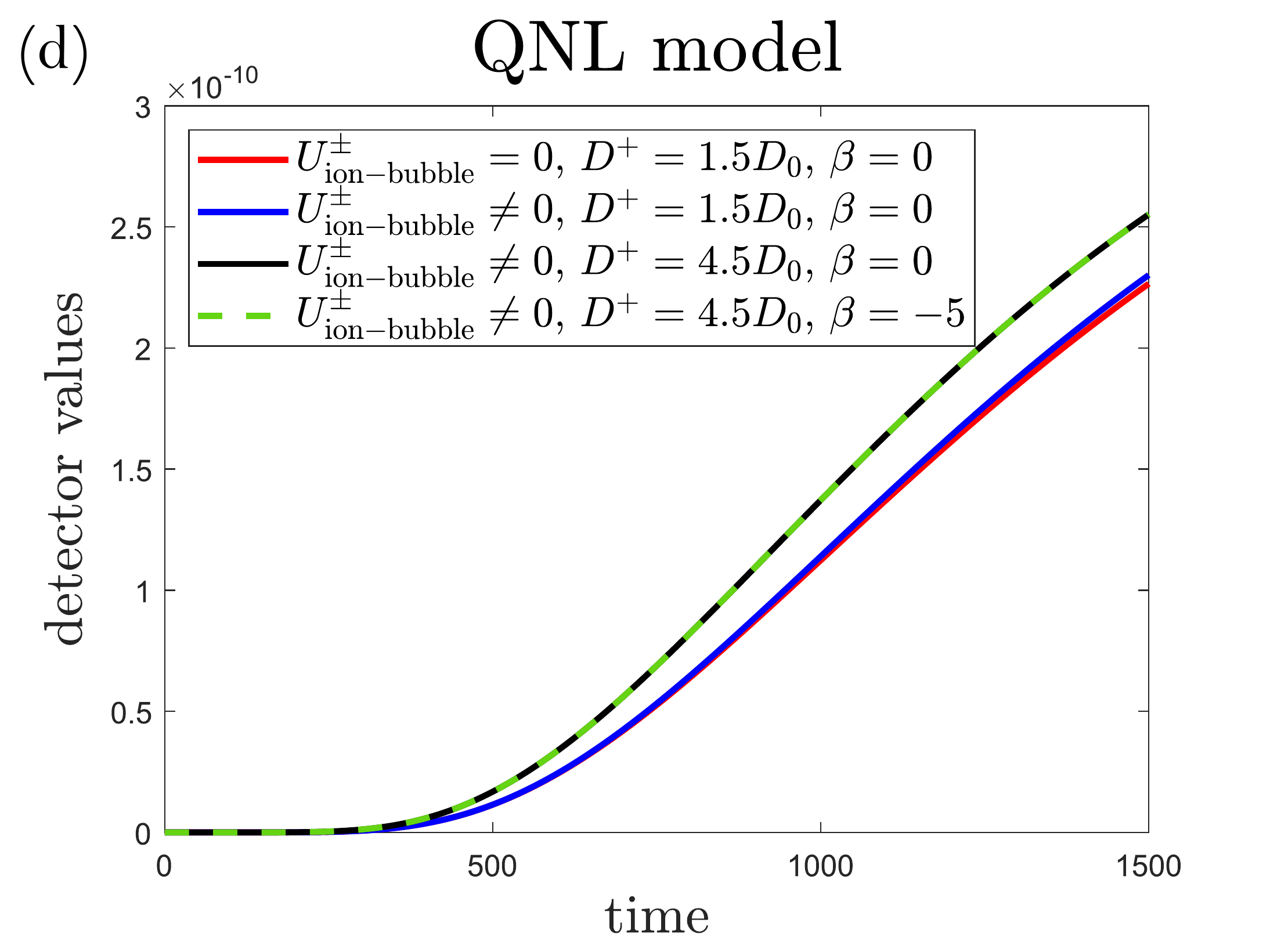}
	\end{minipage}
	\caption{\textit{Values at the detectors of $C^{\varepsilon,+}+C^{\varepsilon,-}$ versus time for different values of $\varepsilon \in \{\infty \,(a), 10^{-4}\, (b), 10^{-5}\, (c)\}$ (where $U^\pm_{i-b} := U^\pm_{\rm{ion-bubble}}$ for simplicity). Panel (d) shows the solution $C$ (multiplied by 2) of the QNL model. The detectors are located at distance $H$ from the bottom. In case \textit{(a)}  {we consider} concentrations $10^4$ times larger  {than $v_0^+$}. The red lines describe the diffusion of the concentrations without the bubble (i.e., $U^\pm_{\rm{ion-bubble}} = 0$). The blue lines describe the behavior of the ions in presence of a bubble and the black lines are obtained using a larger diffusion coefficient. Only for the dashed green lines $\beta = -5 \neq 0$. Grid resolution: $h = 0.04$, $\Delta t=1$ in (a),(b),(d) and $\Delta t = 0.1$ in (c).
	}}
	\label{detector_phi_plot}
\end{figure}
In Fig.~\ref{detector_phi_plot} we report the time evolution of the sum of ion charge densities, $C^{\varepsilon,+} + C^{\varepsilon,-}$, overtaking the bubble and reaching the detector for different values of $\varepsilon \in \{\infty\, (a), 10^{-4} \,(b), 10^{-5}\,(c)\}$. Panel (d) shows the solution {  $C$ (multiplied by a factor 2) } of the QNL model {  \eqref{limit}}, at the detector. 
At $t=0$ a drop of surfactant is put at the bottom of the vessel and is left to diffuse in the solution. At the center of experimental setup a bubble is suspended (see Fig.~\ref{plot_setup} (a)). In the laboratory experiments the total charge of the diffusing species was detected just above the bubble through alternate current conductivity measurements (see Fig.~\ref{plot_setup} (b) and the related discussion). In our computations
the detectors (marked in red, see Fig.~\ref{plot_setup} (a)) are located at distance $H$ from the bottom, (Table~\ref{table_parameters}). 

The red lines in Fig.~\ref{detector_phi_plot} are the solutions of the PNP equations with no bubble (i.e., $\displaystyle U^\pm_{\rm{ion-bubble}} = 0$); the blue lines show the solution of the Eqs.~(\ref{equation_c_U_eps+}-\ref{equation_IC}) obtained using experimental values for the diffusion coefficients for the ions (reported in Table~\ref{table_parameters}); the black lines describe the solution of the Eqs.~(\ref{equation_c_U_eps+}-\ref{equation_IC}) with a larger diffusion coefficient for the cations up to $4.5D_0$, which should mimic the effect of bubble motion. All previous cases consider $\beta = 0$ while for the dashed green lines $\beta=-5$. In panels (b) and (c) we do not see any difference between black and dashed green lines because we consider a volume $v_0^+ = 10^{-6}$ and the product $\beta c$ is negligible. Steric effects become important on the evolution of various species concentrations and on the electrostatic potential for large current densities \cite{steric_effect} and this is the main reason why in almost all previous tests we pose $\beta = 0$. On the contrary, posing $\varepsilon = \infty$ and removing the stiff part of the problem,  {we consider} higher concentrations in panel (a) (i.e., $v_0^+=10^{-2}$), for which the effect of the term $\beta c$ becomes noticeable.  Here again we observe for $\varepsilon = 10^{-4}$ (panel (b)) the behaviour at the detector is similar to the one in panel (a), where anions and cations diffuse independently. We start to see a correlation between the two species choosing, again,  $\varepsilon = 10^{-5}$. 

We want to remark the overshoot was seen only in presence of a vibrating bubble in laboratory experiments. The results reported in Fig.~\ref{detector_phi_plot} suggest a different explanation for the role of surface oscillations with two different diffusion coefficients (for anions and cations). The figure shows that the overshoot is sensitive to the 'effective' (i.e., advection-modulated) diffusion coefficient. Therefore, it is conceivable that even a modest increment of the 'effective' diffusion coefficient $D^+$ (from $1.5D_0$ to  $4.5D_0$) may produce an overshoot at early times.  
The idea that oscillations may re-normalize transport properties in fluid systems is widely supported by different theoretical and experimental papers (for a recent research on the topic see, for instance, \cite{Peraud10829}). However, the results of such simulations are only qualitative, because the potential adopted to describe the bubble is very far from a  {realistic one}. 
\begin{figure}[tb]
	\centering
	\begin{minipage}
		{.49\textwidth}
		\centering
		\includegraphics[width=1\textwidth]{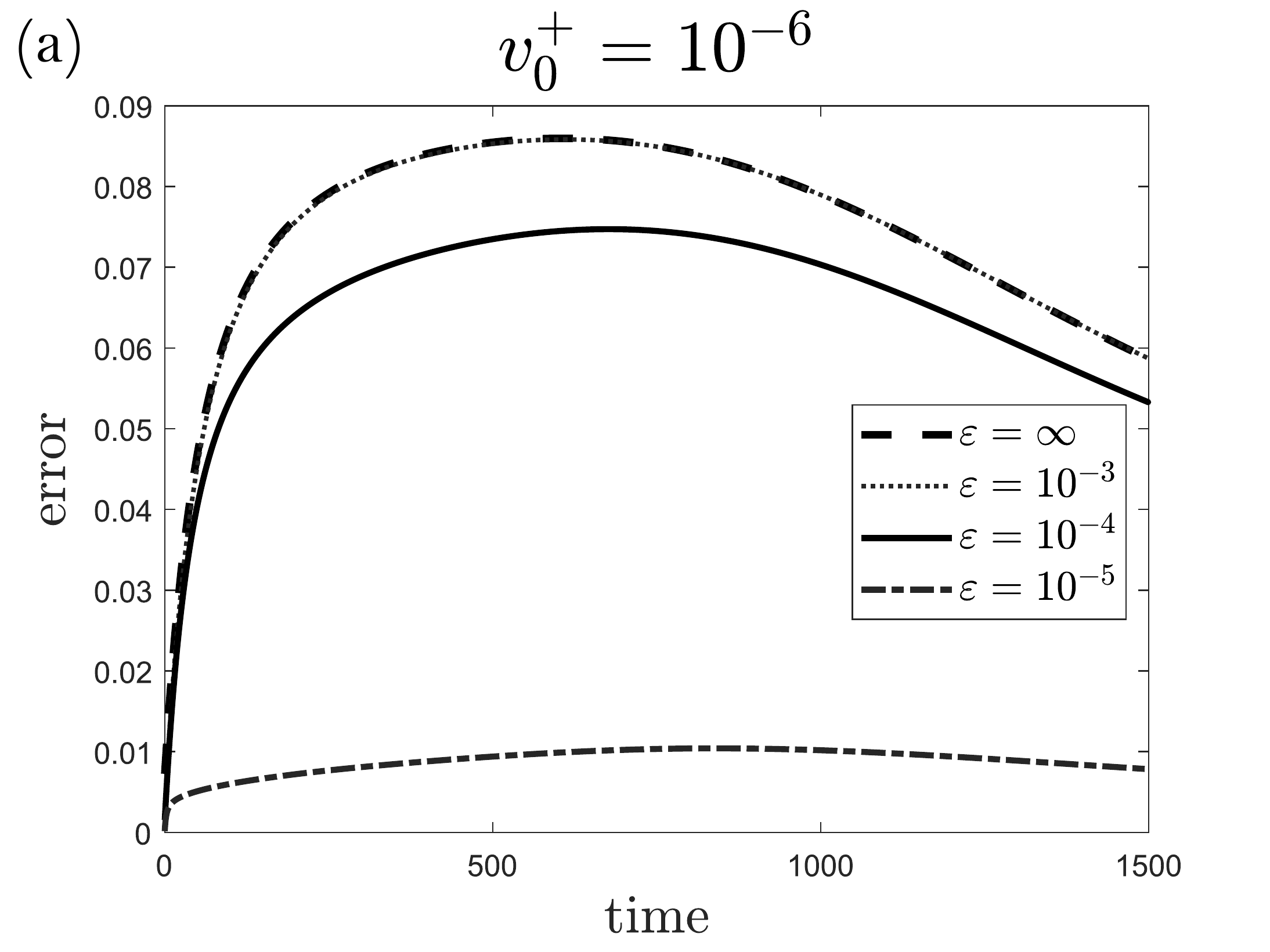}
	\end{minipage}
	\begin{minipage}
		{.49\textwidth}
		\centering
		\includegraphics[width=\textwidth]{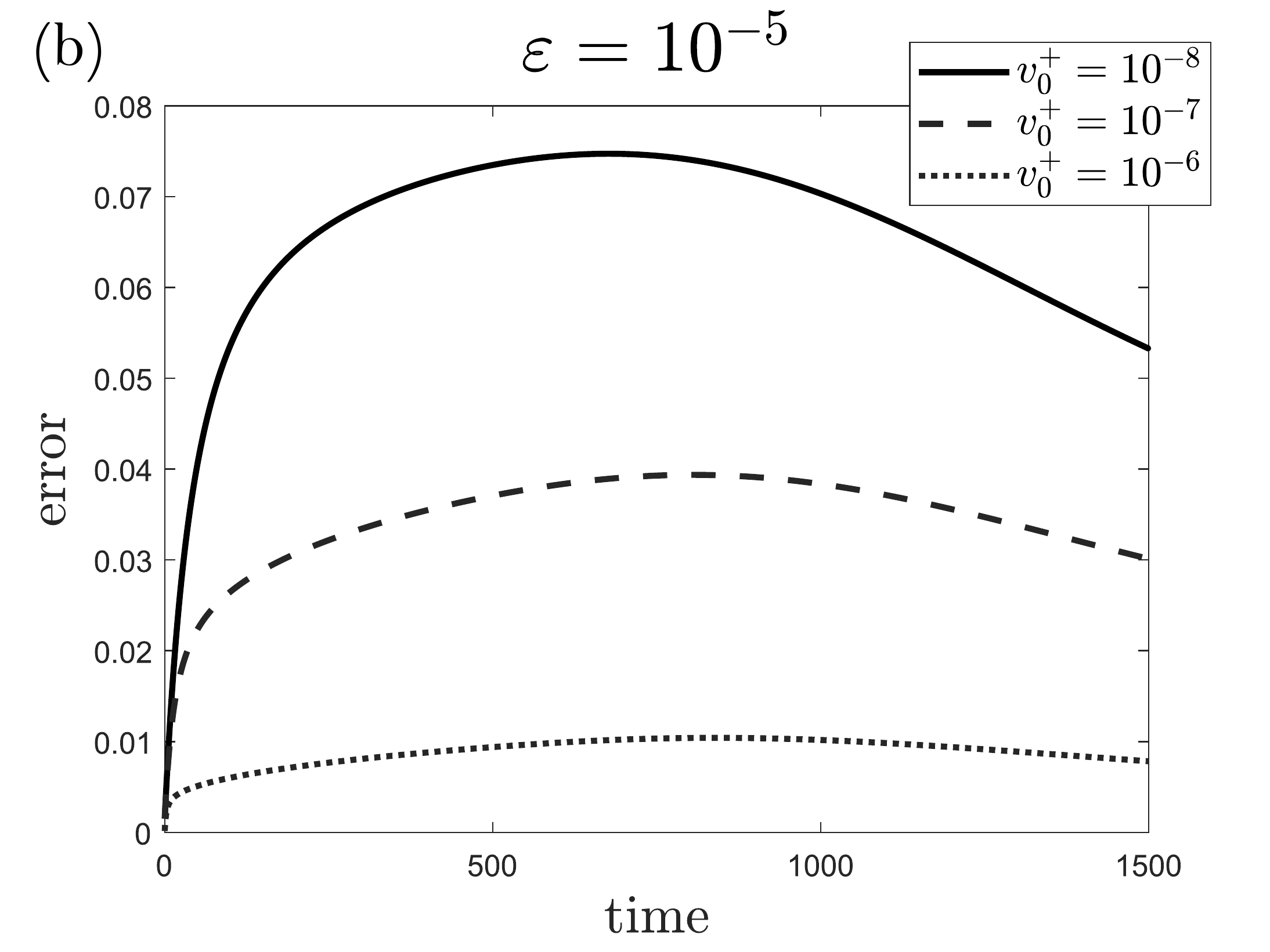}
	\end{minipage}
	\caption{\textit{Relative error in $\mathcal{L}^2$-norm (Eq.~\eqref{relative_error}) between PNP and QNL models as a function of time for different values of $\varepsilon$ (a) and $v_0^+$ (b). Space step $h = 0.04$ and time step $\Delta t = 1$.}}
	\label{plot_error}
\end{figure}

Finally in the last two figures we present a direct comparison between PNP and QNL models. First we show how the relative error in $\mathcal{L}^2$-norm between the two solutions depends on $\varepsilon$, Fig.~\ref{plot_error} (a), and on the total volume $v_0^+$ Fig.~\ref{plot_error} (b). The error is calculated in the whole computational domain as function of time and its expression is the following:
\begin{equation}
	\label{relative_error}
	{ \rm error} = \frac{||0.5(C^{\varepsilon,+} + C^{\varepsilon,-}) - C||_2}{|| C ||_2}
\end{equation}
where $C^{\varepsilon,\pm}$ are solutions of Eqs.~(\ref{QNL-system+}-\ref{QNL-system}) and $C$ is solution of Eq.~\eqref{limit}. In these figures we observe the relative error decreases with $\varepsilon$ (a), as we expected, and again for $\varepsilon = 10^{-5}$ we see a good agreement between the two models. In panel (b) we also see the relative error decreasing when we increase the total volume of the particles. Because of the non-linearity of the Coulomb term in the equations, its effect is stronger for larger concentrations. 

\begin{figure}[tb]
	\centering
	\begin{minipage}
		{.32\textwidth}
		\centering
		\includegraphics[width=\textwidth]{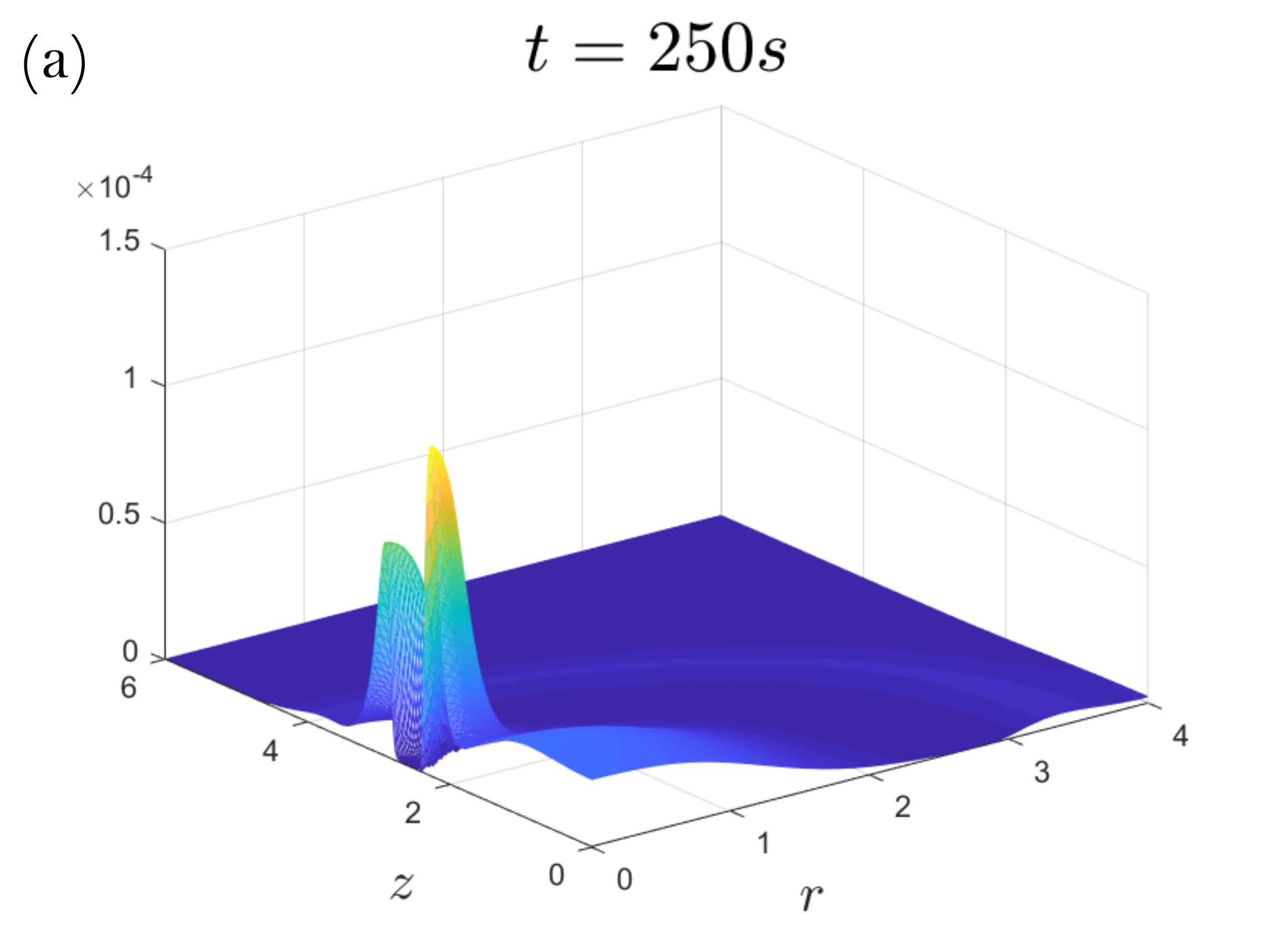}
	\end{minipage}
	\begin{minipage}
		{.32\textwidth}
		\centering
		\includegraphics[width=\textwidth]{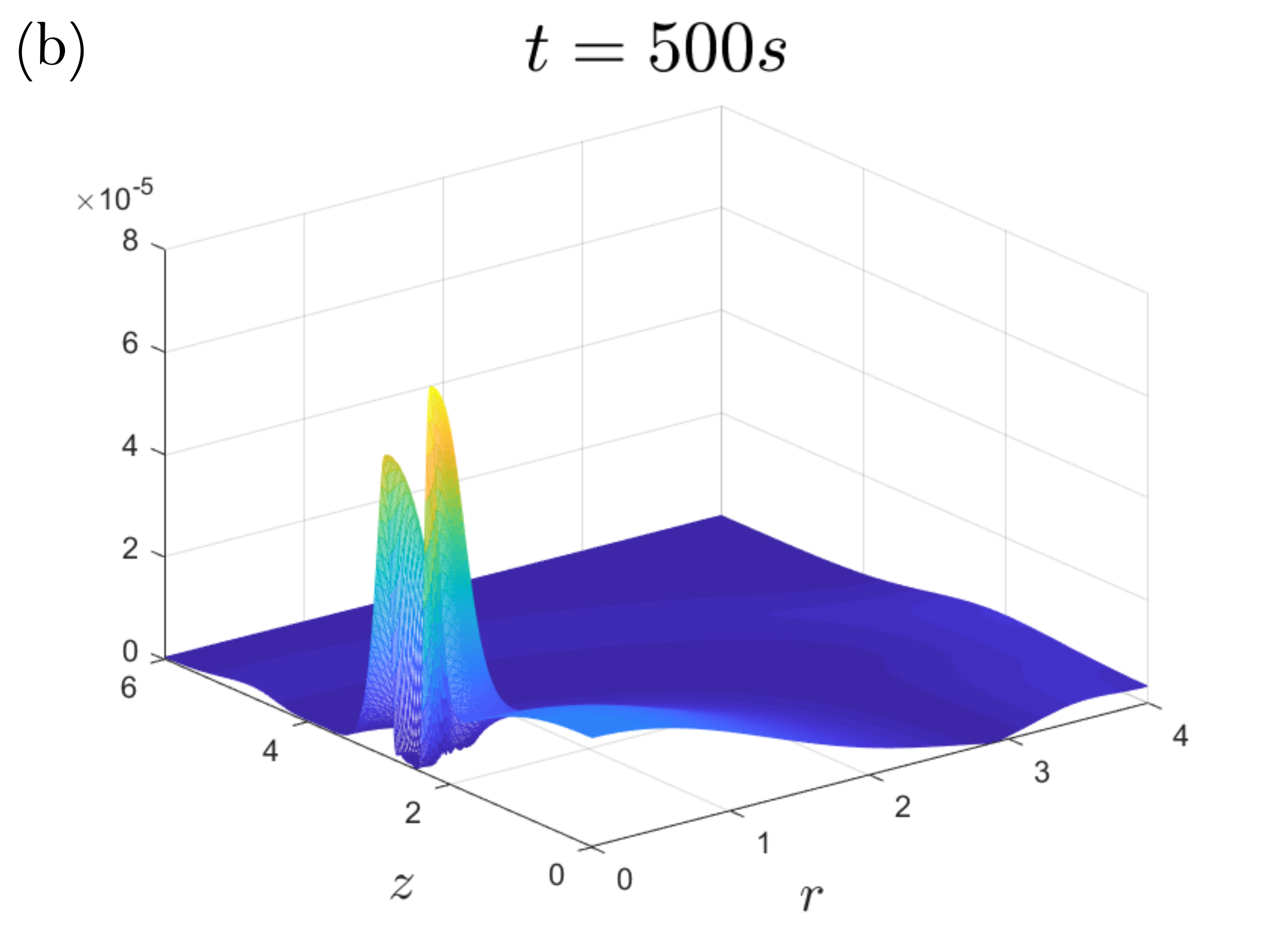}
	\end{minipage}
	\begin{minipage}
		{.32\textwidth}
		\centering
		\includegraphics[width=\textwidth]{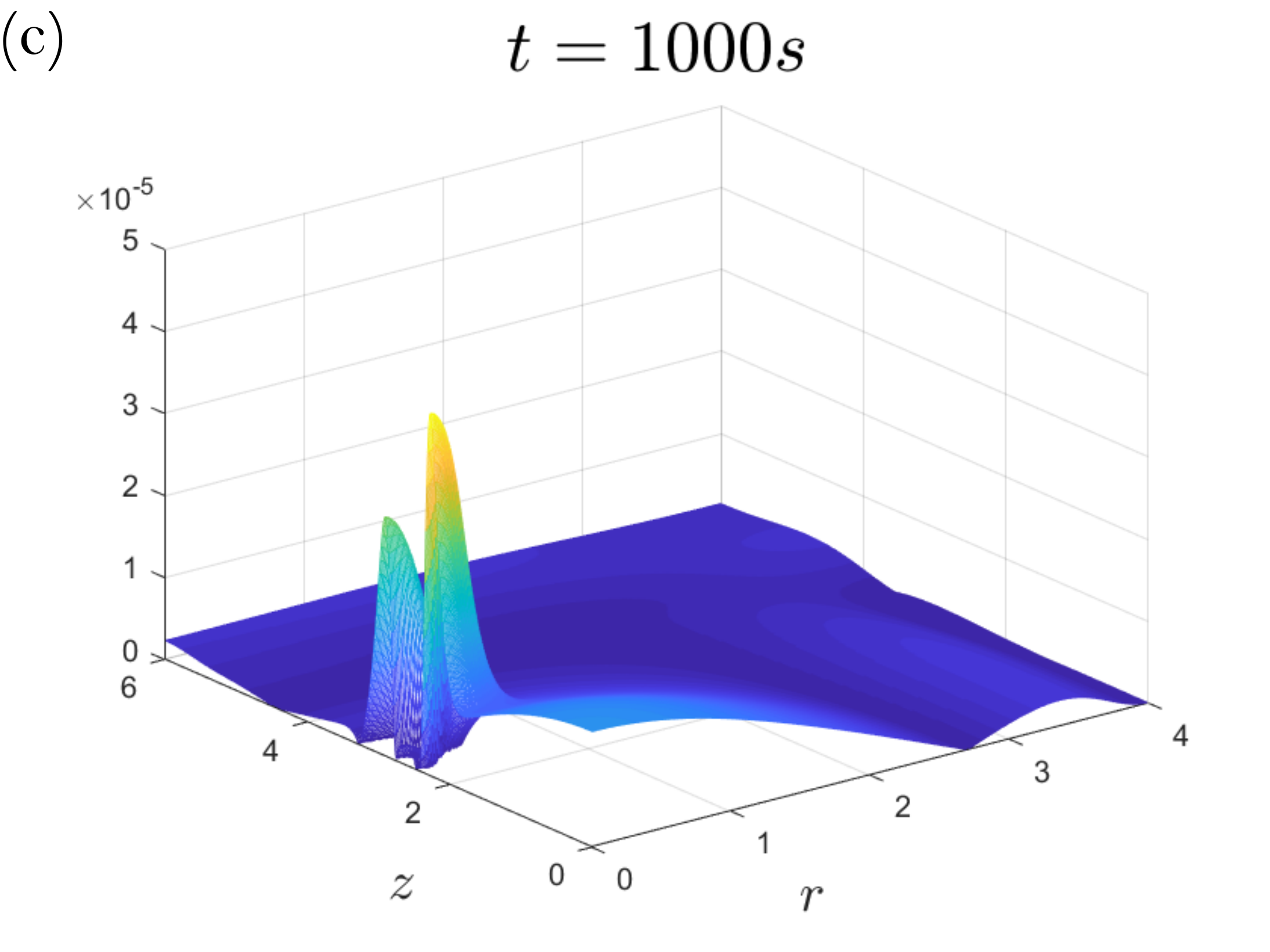}
	\end{minipage}
	\caption{\textit{Agreement between models. In these plots we show the error between the solutions of PNP (Eqs.~(\ref{QNL-system+}-\ref{QNL-system})) and QNL models (Eq.~\eqref{limit}) (i.e., we plot $|(0.5(C^{\varepsilon,+}+C^{\varepsilon,-})-C)|/\int_\Omega Cd\textbf{r}$) for different times, $t \in\{t=250s\, (a), t=500\, (b) t=1000s\, (c)\}$. After large time the difference is non negligible only near the bubble surface that means quasi-neutral limit is a very good approximation far from the action of the bubble. 
	In these tests $\varepsilon = 10^{-6}$, spatial step is $h = 0.0133$ and time steps $\Delta t = 1$ for QNL model and $\Delta t = 0.005$ for PNP model. }}
	\label{QNL_comparison_plots}
\end{figure}

In Fig.~\ref{QNL_comparison_plots} we show the quantity $|(0.5(C^{\varepsilon,+}+C^{\varepsilon,-})-C)|/\int_\Omega Cd\textbf{r}$ for three times \\$t\in\{250s \, (a), 500s\, (b), 1000s\, (c)\}$. We notice the difference is almost zero far from the action of the potentials which means quasi-neutrality is a good approximation for the PNP model, except near the bubble. The approximation would be much better with more realistic values of $\varepsilon$.
\section{Conclusions}
In this paper we use a simple Poisson-Nernst-Planck model to study the correlated ion diffusion in presence of a trap described by a potential. A coupled set of equations describing cations and anions motion is solved numerically. The trap (an air bubble in our experiments) is modeled by a short-range repulsive potential for the cations and by a combination of an attractive and repulsive potential acting on the anions. Such different kinds of potentials mimic the different chemical nature of the considered ions: anions bring a large hydrophobic tail which pushes them toward the bubble surface. On the contrary, cations are small charged particles which prefer to stay inside water and far from the bubble surface.

The ADI method we propose, with the extrapolation technique, is an improvement of the standard one. We can easily see the method we describe has the same accuracy of the standard second-ordered one but it is more efficient because it uses less  CPU time for the same error. The method is based on the extrapolation of the concentration used to compute the potential $\phi$ in an optimal way. The convergence rate is calculated numerically and confirms the second order accuracy. 

A simplified model PNP system has been introduced, which makes it easier to test new numerical schemes. In particular, we propose a semi-implicit scheme which is more stable for $\Delta t$ larger than $\varepsilon$ but it is not exactly conservative. The conservation mass error depends on the time discretization, being first order with $\Delta t$, as expected.

Numerical experiments show a non-monotonic behavior of the total ionic density detected past the bubble, in qualitative agreement with laboratory experiments but only for unrealistic values of Debye length and bubble thickness. 

The real problem is multiscale in space since the range of the ion-bubble (few nanometers), and ion-ion potentials (tens of nanometers) are orders of magnitude smaller than the size of the bubble.

 {Furthermore}, laboratory experiments show non monotonic behavior only in presence of oscillating bubble, with oscillation resonance frequency of few hundred Hz. Diffusion time scale is approximately one hour, therefore this is a multiscale problem also in time.

Because of the space and time multiscale nature of the problem, direct numerical simulation using the model presented in the paper is not feasible.\\
In order to perform a more quantitative simulation, we are developing a multiscale model based on a suitable boundary condition on the bubble surface that describes the effect of the (thin) bubble which overcomes the multiple scales in space, together with a homogenization technique in time which separates fast (oscillations) and slow (diffusion) time scales. Furthermore, even with realistic values of Debye length, charge neutrality may break down near the bubble surface, as suggested by several experiments (see, for instance, \cite{LEVY2020162,101137140968082}), which poses additional challenges both at modeling and computational levels. 

Whether or not the overshoot seen in laboratory experiments is a sign of local electroneutrality breakdown and other approximations are left to future research. In principle, this hypothesis is reasonable because the investigated system considerably deviates from the one of classical Coulomb plasmas (attractive bubble-anions and anions-anions interactions are missing in classical models of ideal plasma). At this moment, however, this is a suggestive hypothesis that requires lot of new models and simulations in order to be proved.

\section*{Acknowledgments}
A.R., G.L. and A.G. thank the University of Catania (Piano della Ricerca di Ateneo 2016-2018) for partial financial support. 

G.R. and C.A. thank ITN-ETN Horizon 2020 Project ModCompShock, Modeling and Computation on Shocks and Interfaces, Project Reference 642768, and the Italian Ministry of Instruction, University and Research (MIUR) to support this research with funds coming from PRIN Project 2017 (No.2017KKJP4X entitled ''Innovative numerical methods for evolutionary partial differential equations and applications''. 

\newpage

\begin{appendix}
\label{appendix_chapter}
\subsection{Derivation of the diffusion equation for binary electrolyte solution.}
\label{section_derivation}
A brief sketch of the derivation of Eqs.~(\ref{equation_flux}-\ref{equation_definition_flux+}) is as follows. The average velocity $v^i$ of the $i$-th ion species is proportional to the average of the sum of all the forces acting on the moving ions. In 1D the system is:
\begin{equation}
	\displaystyle v^i=-u^i\frac{\partial}{\partial x}\left(\mu^i+V^i\right)
	\label{equation_1D}
\end{equation} 
where $u^i$ is the ion mobility (that may depend upon the ion concentration, see below) and $\displaystyle -\frac{\partial}{\partial x}(\mu^i+V^i)$ is the gradient of the chemical potential (defined as the derivative of the energy with respect to the concentration) partitioned, as usual, into an entropic ($\mu^i$) and an interaction ($V^i$) contribution. The entropic component of the chemical potential depends on the specific form of entropy we adopt. In the simplest case of Boltzmann entropy (valid for dilute solutions), the entropy of the moving species takes the simple form:  $S_{mix}=\ -k_Bc^i\log{c^i}$, thus: {$\displaystyle \mu^i=k_BT\frac{\partial}{\partial c^i}\left(c^i\log{c^i}\right)=k_BT(\log{c^i}+1)$}. 

For concentrated solutions and considering the correct form of entropy, we should take into account the reduced number of combinations using: 
\[\displaystyle S_{mix}=\ -k_B\left(\Sigma_i c^i\log{c^i}+\left(1-\Sigma_i c^i\right)\log{\left(1-\Sigma_i c^i\right)}\right) \]
where the last term describes the mixing entropy of solvent molecules (the concentration of which is: $1-\Sigma_i c^i$). As mentioned in the main text, for dilute solutions $c^i \ll 1$ and therefore we neglet the term $(1-\Sigma_i c^i)\,\log(1-\Sigma_i c^i)$ with respect to $c^i \log\,c^i$.

The structure of the interaction term $V^i$ can be very complicated. In the case of simple point-like ions it contains only electrostatic components, while in the present study both electrostatic and hydrophobic contributions are presented at the same time. Performing the derivatives and plugging them into Eq.~\eqref{equation_1D} yields in the low concentrations limit:
\begin{eqnarray}
	\displaystyle	v^i=-\frac{u^ik_BT}{c^i}\left(\frac{\partial c^i}{\partial x} +\frac{c^i}{k_BT}\frac{\partial V^i}{\partial x} \right)
\end{eqnarray}
Since the ion flux $J^i$ is defined as: $J^i= c^iv^i$, we immediately get:
\begin{equation}
	J^i = -D^i\left(\frac{\partial c^i}{\partial x}+\frac{c^i}{k_BT}\frac{\partial V^i}{\partial x}\right)
	\label{equation_derivation_flux}
\end{equation}
where $D^i\equiv u^ik_BT$ in the hypothesis of constant (concentration-independent) mobility. Eq.~\eqref{equation_derivation_flux} is identical to Eqs.~\eqref{equation_definition_flux+} of the main text (where $i=+\ or \ -$). 

In general, however, $u^i$ depends on the local concentrations $c^i$ of the diffusant species (see, e.g., \cite{Rashidnia:02}). 
\subsection{Dimensional Analysis}
\label{section_dimensionless}
Here we rewrite the equations using units which are more suitable for the problem. 
Starting from the Eqs.~(\ref{equation_cphi+}-\ref{equation_cphi-})  we have
\begin{eqnarray}	
		\displaystyle  \tilde{\partial}_t c^\pm &=& \tilde{D}^\pm {\Delta}  c^\pm + \tilde{\chi}_\pm \tilde{\nabla} \cdot \left(c^\pm \tilde{\nabla}\tilde{V}^\pm\right)\\
		\displaystyle & =& \tilde{D}^\pm\left( {\Delta}  c^\pm + \tilde{\nabla} \cdot \left(c^\pm\tilde{\nabla}{\tilde{U}^\pm}\right) \right)
\end{eqnarray}
where we denote by $\tilde{\cdot}$ dimensional quantities expressed in SI system units, $\quad \tilde{\chi}_\pm = {\tilde{D}_\pm}/({k_BT})$ and we define $\displaystyle \tilde{U}^\pm :=\frac{\tilde{V}^\pm}{k_BT}$. Hence the potentials $\tilde{U}^\pm$ take the following form from the expressions in Eq.~\eqref{equation_definition_ext}
\begin{eqnarray}	
		\displaystyle \tilde{U}^+ &= &\tilde{U}_{\rm{ion-bubble}}^+ + \frac{q\tilde{\varphi}}{k_BT} \\
		\displaystyle \tilde{U}^- &= &\tilde{U}_{\rm{ion-bubble}}^- - \frac{q\tilde{\varphi}}{k_BT} - \frac{\beta}{k_BT} c^-
\end{eqnarray}
where $q $ is the (positive) electron charge and $\displaystyle \tilde{U}_{\rm{ion-bubble}}^\pm := \frac{\tilde{V}_{\rm{ion-bubble}}^\pm}{k_BT}$. \\
We rewrite the Poisson Eq.~\eqref{equation_poisson_phi} for the electrostatic potential $\tilde{\varphi}$ as follows:
\begin{equation}
	\label{equation_phi}
-\epsilon_0\epsilon_r\Delta \tilde{\varphi} = q(n^+-n^-) 
\quad {\left[\rm{Coulomb}/m^3\right]}
\end{equation}
where $\epsilon_0 $ is the vacuum permittivity, $\epsilon_r $ is the  relative permittivity and
\begin{equation}
	n^\pm = \frac{c^\pm N_A \rho^\pm}{\tilde{m}^\pm }, \qquad \left[\# \rm{ions}/m^3\right]
	\label{expr_n_pm}
\end{equation}
with $N_A $ the Avogadro's number, $\tilde{m}^\pm$ the molar mass of ions 
(Kg/mol) and $ \rho^\pm$ their mass densities  {(Kg/m$^3$)}. Multipling Eq.~\eqref{equation_phi} by $q/(\epsilon_0 \epsilon_r)$ and replacing $n^\pm$ with Eq.~\eqref{expr_n_pm}, we obtain
\begin{eqnarray}
		\displaystyle -  {\Delta} \tilde{V}^\varepsilon &= &\frac{q^2 N_A }{\epsilon_0\epsilon_r}\left(\frac{c^+\rho^+}{ \tilde{m}^ +}-\frac{c^-\rho^-}{ \tilde{m}^ -}\right)
		\label{phiNond}
\end{eqnarray}
where $\tilde{V}^\varepsilon := q\tilde{\varphi}$. With the assumptions $\rho:= \rho^+ = \rho^-$ 
and ${\tilde{m}}^ \pm = m_0 m^\pm$ Eq.~\eqref{phiNond} becomes
\begin{eqnarray}	
		\displaystyle -  {\Delta}   \tilde{V}^\varepsilon &=& q^2\frac{N_A\rho }{\epsilon_0\epsilon_r m_0}\left(\frac{c^+}{m^+}-\frac{c^-}{m^-}\right)  \\
		\displaystyle  &=& \tilde{K}\left(\frac{c^+}{m^+}-\frac{c^-}{m^-}\right)
	\label{phiNond2}
\end{eqnarray}
where $\displaystyle \tilde{K}=\frac{q^2N_A \rho}{\epsilon_0\epsilon_r m_0}$. 
Now we divide Eq.~\eqref{phiNond2} by $k_BT$, obtaining 
\begin{eqnarray}
	\label{Kformulation2}
	\displaystyle -  {\Delta} U^\varepsilon &=& K\left(\frac{c^+}{m^+}-\frac{c^-}{m^-}\right) \qquad \left[\#k_BT /m^2\right]
\end{eqnarray}
where  $\displaystyle U^\varepsilon := \frac{\tilde{V}^\varepsilon}{k_BT}$ and $ \displaystyle K = \frac{\tilde{K}}{k_BT} = \frac{q^2 N_A\rho}{\epsilon_0\epsilon_r k_BT m_0}$.

The \textit{Debye length} is defined as $\lambda_D = \left(Kc^+/m^+\right)^{-1/2}$. In our model $c^+ \approx 10^{-6}$, which gives $\lambda_D$ of the order of nanometers. The last quantity we define is $\varepsilon = \lambda_D \sqrt{c^+/m^+} = K^{-1/2}$ and the Eq.~\eqref{Kformulation2} becomes
\begin{equation}
	-\varepsilon^2 {\Delta}U^\varepsilon=\frac{c^+}{m^+}\ -\ \frac{c^-}{m^-}	
\end{equation}
using the values reported in Table~\ref{table_parameters} its numerical value is $\varepsilon \approx 1.36\times 10^{-8}$.

\subsection{Algorithm for second order ADI method for PNP model}
\label{section_ADI}
Here we describe the steps of the ADI method applied to PNP model. We first compute the concentration in the first half step $n+1/2$ with an extrapolation technique to compute the potential at the same time step. After that we solve the Eqs.~(\ref{equation_c_U_eps+}-\ref{equation_lap_Ueps}) for $c^{n+1/2}$ in the first half step $n+1/2$ and for $c^{n+1}$ in the step $n+1$ with $c^{\pm,n} := c^{\varepsilon,\pm}(n\Delta t)$ and $\Delta t$ the time step.

Given $c^{\pm,n}$ and $c^{\pm,n-1}$ we find $c^{\pm,n+1}$ as follows: 
\begin{itemize} 
\item compute $c^{\pm,n+1/2}$ from the previous two time steps
\[ c^{\pm,n+1/2} = \frac{3}{2}c^{\pm,n} - \frac{1}{2}c^{\pm,n-1} \] 
\item compute the Coulomb potential by solving the discrete Poisson equation in  {the} grid $\Omega_h$
 
{\[
	 - \mathcal{L} \,U^{\varepsilon,n}  ={\varepsilon^{-2}}\left(
	\frac{c^{+,n+1/2}}{m^+}-\frac{c^{-,n+1/2}}{m^-}\right) 
\]
}
\item compute the total potential
\begin{eqnarray}
	\nonumber
	 {\displaystyle U^{\pm,n} = U_{\rm{ion-bubble}}^\pm \pm U^{\varepsilon,n} \quad  \text{ in } \Omega_h}
\end{eqnarray}
\item compute $c^{\pm,n+1/2}$ (implicit in $r$, explicit in $z$)
\begin{eqnarray}
	\nonumber
	\displaystyle \frac{c^{\pm,n+1/2}-c^{\pm,n}}{\Delta t} & = \frac{ D^\pm }{2} \left( \mathcal{L}_r  c^{\pm,n+1/2} +  \mathcal{D}_r \cdot \left(c^{\pm,n+1/2} \mathcal{D}_r   U^{\pm,n} \right) +  \mathcal{L}_z  c^{\pm,n} \right)  \\ \nonumber & \displaystyle   {+ \frac{ D^\pm }{2} \left(\mathcal{D}_ z\cdot\left(c^{\pm,n} \mathcal{D}_z  U^{\pm,n} \right)\right) \quad \text{ in } \Omega_h }
\end{eqnarray}
\item compute $c^{\pm,n+1}$ (implicit in $z$, explicit in $r$)
\begin{eqnarray}
	\nonumber
	 \displaystyle \frac{c^{\pm,n+1}-c^{\pm,n+1/2}}{\Delta t} &= \frac{D^\pm}{2} \left( \mathcal{L}_z  c^{\pm,n+1} + \mathcal{D}_ z\cdot\left(c^{\pm,n+1} \mathcal{D}_ z  U^{\pm,n} \right) +  \mathcal{L}_r  c^{\pm,n+1/2}\right)  \\ 
	\nonumber &  { \displaystyle  + \frac{D^\pm}{2} \left(\mathcal{D}_r \cdot\left(c^{\pm,n+1/2} \mathcal{D}_r   U^{\pm,n} \right)\right) \quad \text{ in } \Omega_h} 
\end{eqnarray}
\end{itemize}

$ \displaystyle \mathcal{D}_\alpha$ and $\displaystyle  \mathcal{L}_\alpha $, with $\alpha=r,z$, are the discrete operators for gradient (or divergence) and Laplacian in $r$ and $z$ direction respectively with $\displaystyle  \mathcal{D}_r ,  \mathcal{L}_r \in  \mathbb{R}^{N_r\times N_r}$ and $\displaystyle \mathcal{D}_z, \mathcal{L}_z \in \mathbb{R}^{N_z\times N_z}$, while $\mathcal{L}$ is the discrete operator for the Laplacian in both directions ($r$ and $z$) with $\mathcal{L}\in \mathbb{R}^{N_r N_z\times N_r N_z}$, corresponding to zero Neumann condition.

Now we define the ADI method for one species model in QNL 
for $\displaystyle C:=\frac{c^{\varepsilon,\pm}}{m^\pm}$
\begin{eqnarray}
\nonumber
& \displaystyle \frac{C^{n+1/2}-C^{n}}{\Delta t/2} = { D_{\rm{eff}} } \left( \mathcal{L}_r  C^{n+1/2} + \mathcal{L}_z  C^{n} +  \mathcal{D}_r \cdot\left(C^{n+1/2} \mathcal{D}_r  W \right)\right) \\ 
\nonumber & + { D_{\rm{eff}} } \left( \mathcal{D}_z\cdot\left(C^{n} \mathcal{D}_ z  W  \right)\right) \quad  { \text{ in } \Omega_h }\\ 
& \displaystyle \frac{C^{n+1}-C^{n+1/2}}{\Delta t/2} = D_{\rm{eff}} \left(\mathcal{L}_z  C^{n+1}+ \mathcal{L}_r  C^{n+1/2} + \mathcal{D}_ z\cdot\left(C^{n+1} \mathcal{D}_ z  W  \right)\right)  \\ \nonumber &  {+ D_{\rm{eff}} \left(\mathcal{D}_r \cdot\left(C^{n+1/2} \mathcal{D}_r  W  \right)\right) \quad \text{ in } \Omega_h }
\end{eqnarray}
where for simplicity of notation $ W := U_{\rm{ion-bubble}} =  (U^+_{\rm{ion-bubble}}+ U^-_{\rm{ion-bubble}})/2$. 

Considering the QNL approximation we solve only linear systems of dimension $N_r\times N_r$ and $N_z\times N_z$ avoiding the linear system of dimension ${N_r N_z}\times{N_r N_z}$ (coming from the Poisson equation), drastically reducing the computational cost. \bigskip

The method is implemented in Matlab on a Dell Inspiron 13-5379, 8th Generation Intel Core i7, 16GB RAM.

\subsection{Richardson extrapolation}
\label{app:Richardson}
Here we show how to use Richardson extrapolation to estimate the error of a given method, once the order $p$ of the method is known. 
Let us denote by $u_{\rm exa}$ the exact solution of a problem, by $u(h)$ the solution that depends on a discretization parameter $h$, and by $u(0)$ the limit solution obtained as $h\to0$. Assume that the solution is regular, and depends smoothly on the parameter $h$. 
One has:
\begin{align*}
    u(h) & =  u(0) + C h^p + o(h^p)\\
    u(qh) & =  u(0) + C (qh)^p + o(h^p)
\end{align*}
Subtracting the second relation from the first, one has:
\[
    u(h)-u(qh) = Ch^p(1-q^p)+o(h^p)
\]
from which it follows
\[
    Ch^p = \frac{u(h)-u(qh)}{1-q^p}+o(h^p)
    = u(h)-u(0) + o(h^p)
\]
For $p=2$ and $q=1/2$, one has 
\[
    u(h)-u(0) = \frac{4}{3}(u(h)-u(h/2)) + o(h^p)
\]
Notice that convergence of order $p$ requires the additional assumption that $u(0) = u_{\rm exa}$, i.e.\ that the method is convergent. Once convergence is assessed, then extrapolation can be adopted to estimate the error. 
\end{appendix}


\end{document}